\documentclass[aps,pra,reprint]{revtex4-1}

\usepackage[hyperindex,breaklinks]{hyperref}
\usepackage{graphicx}
\usepackage{dcolumn}
\usepackage{bm}
\usepackage{multirow}
\usepackage{upgreek}
\usepackage{braket}
\usepackage{bm}        
\usepackage{amssymb}   
\usepackage{amsmath}
\usepackage{subfigure}
\usepackage{epstopdf}
\usepackage{comment}
\usepackage{float}
\usepackage{xcolor}
\usepackage{hyperref}

\def\D{\Delta}

\begin{document}

\title{Mechanism of stimulated Hawking radiation in a laboratory Bose-Einstein condensate}

\author{Yi-Hsieh Wang$^{1,2}$}
\author{Ted Jacobson$^{3}$}
\author{Mark Edwards$^{1,4}$}
\author{Charles W. Clark$^{1,2}$}
\affiliation{$^1$Joint Quantum Institute, National Institute of Standards and Technology and the University of Maryland, College Park, Maryland 20742,
\\$^2$Chemical Physics Program, University of Maryland, College Park, Maryland 20742,
\\$^3$Department of Physics and Maryland Center for Fundamental Physics, University of Maryland, College Park, Maryland, 20742,
\\$^4$Physics Department, Georgia Southern University, Statesboro, GA 30460}

\begin{abstract}
We model a sonic black hole analog in a quasi one-dimensional Bose-Einstein condensate, using a Gross-Pitaevskii equation matching
the configuration of a recent experiment by Steinhauer [Nat. Phys. {\bf 10}, 864 (2014)]. The model
agrees well with important features of the experimental observations, 
demonstrating their hydrodynamic nature.
We find that a zero-frequency bow wave is generated at the inner (white hole)
horizon, which grows in proportion to the square of the background condensate density.
The relative motion of the black and white hole horizons produces a Doppler shift of the bow wave at the black hole, where it stimulates 
the emission of monochromatic Hawking radiation. The mechanism is confirmed using 
temporal and spatial windowed Fourier spectra of the condensate. 
Mean field behavior similar to that in the experiment can thus be fully explained without the presence of self-amplifying  Hawking radiation. 

\end{abstract}

\maketitle

\section{Introduction and Summary}
\label{sec:intro}
Hawking radiation \cite{hawking1, hawking2} is a pair creation process, resulting 
from a vacuum instability of quantum fields at a black hole horizon.
The radiation is thermal, with temperature, $T = 62 \, \mathrm{ nK }\, M_{\odot} / M$, for a spherical black hole of mass $M$, where $M_{\odot}$ is the mass of the Sun. Such low temperature Hawking radiation (HR) from solar mass or larger black holes will likely never be observed. However, sonic analogs of HR  can exist in hydrodynamic systems
possessing a ``sonic horizon," where the flow transitions from subsonic to supersonic \cite{unruh1981}.
In such systems, the Hawking temperature is proportional to Planck's constant times
the velocity gradient at the horizon. To produce observable {\it quantum} HR, the system must not
be much warmer than this temperature, and for this reason Bose-Einstein condensates (BECs) 
are a natural candidate \cite{ 
robertson2012,
PhysRevD.85.024021,PhysRevA.80.043601, Gardiner2007,PhysRevLett.113.090405,PhysRevLett.85.4643}.  
In a sonic analog, the experimenter has access to the regions both inside and outside of the horizon, thus enabling measurements of correlations that could exhibit quantum entanglement of Hawking quanta and their interior partners \cite{Balbinot:2007de,PhysRevA.80.043603, PhysRevA.85.013621,Boiron2015,Steinhauer:2015saa,
Steinhauer:2015ava}.

\begin{figure*}[htb]
\centering
\includegraphics[width=7.2in]{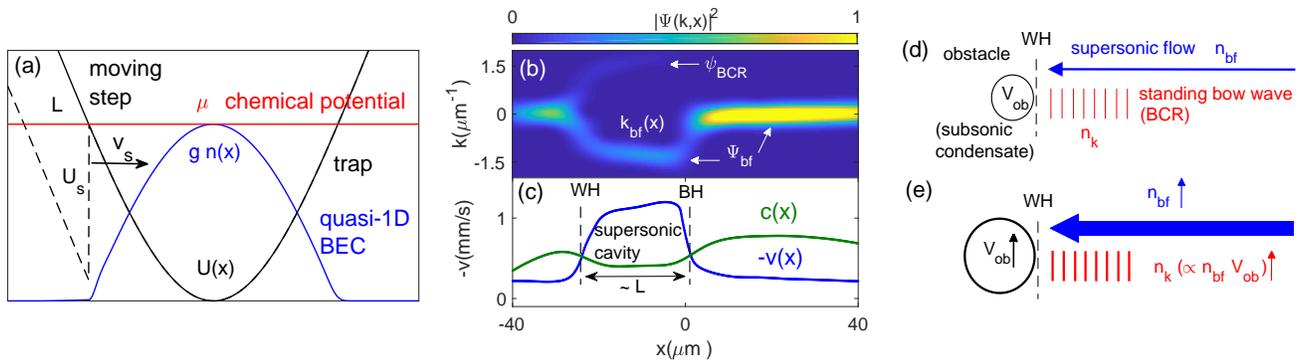}
\caption{\label{fig:exp}  
(a) Potential-step structure at the start of its sweep through the BEC with number density $n(x)$, chemical potential $\mu$, and confining potential $U(x)$; (b) windowed wavevector spectrum of a BEC during the sweep,
showing an accelerated flow
generated by the moving step, edge at $x=0$; the peak location is denoted by $k_{\rm bf}(x)$; (c) flow speed in the step frame
$-v(x)$ and speed of sound $c(x)$, which (in a globally stationary flow)
are equal at the horizons, WH and BH; (d) illustration of a standing wave (BCR) against an obstacle (subsonic BEC) near the WH; (e) growth of the standing wave amplitude ($n_k$) resulting from the increase of background density ($n_{\rm bf}$) and obstacle strength ($V_{\rm ob}$).}
\end{figure*}

In 2014, Steinhauer \cite{nphys3104} reported on a remarkable 
experiment that implements a sonic analog black hole 
in a needle--shaped BEC of $^{87}$Rb. The BEC is swept by a negative 
step potential with the energy equivalent of a few nK. 
The sweep generates a low-density ``cavity" of supersonic flow in the interior of the BEC,
bounded by black-hole (BH) and white-hole (WH) horizons [Figs.~\ref{fig:exp}(a)-(c)].  
Steinhauer observed exponential growth of a standing wave between the horizons, 
and measured the density-density, two-point, connected correlation function, 
which displayed correlations between points within the cavity, as well as between a point in the cavity and a point
outside the BH. He interpreted the latter as a signal of Hawking radiation phonons correlated with their 
partners behind the horizon, and he interpreted the growing standing wave and internal correlation
function as evidence for self-amplifying Hawking radiation.

In a flow with a supersonic 
cavity bounded by BH and WH horizons, 
Hawking radiation can be self-amplifying
if the phonon group velocity 
becomes supersonic at high wavenumbers\cite{jacobson1999}. 
The negative energy partner of a Hawking phonon is trapped in the cavity, 
so that the parent state for subsequent Hawking radiation is no longer the 
vacuum, but instead is an excited state. This produces stimulated
emission of Hawking radiation, which amplifies the trapped negative energy
mode. The repetition of this process leads to exponential growth of the negative energy
mode and the associated Hawking emission. This is called the 
``black hole laser" mechanism\cite{jacobson1999}. 
The behavior of this lasing mechanism, and its potential role
in laboratory realizations of Hawking radiation
has been extensively investigated\cite{Coutant:2009cu,parentani2010}. 
It could enhance the signal of Hawking radiation but, 
even if the initial trigger for the lasing were spontaneous emission of Hawking
radiation, the amplified signal, once it had grown significantly, would be 
a coherent state of phonons which would be difficult to distinguish from 
a classical wave. Moreover, a lasing mode could be excited by a classical seed.

In the work reported here, we have investigated the dynamics of the condensate
of the experiment of Ref. \cite{nphys3104}, using primarily the 
one dimensional, time-dependent Gross-Pitaevskii (GP) equation.
(We used the three-dimensional GP equation only to check that the qualitative 
features of the dynamics are the same as for the one-dimensional case.)
The GP equation is a nonlinear Schrodinger equation, approximating the behavior of the 
expectation value of the many body field operator, which captures the classical,
hydrodynamic aspects of the BEC, as well as interference phenomena. 
Effects of quantum fluctuations can 
be treated approximately
by adding to the GP wavefunction an initial distribution of 
random fluctuations with the Gaussian statistics of the zero point fluctuations, and averaging
over an ensemble of realizations. That is called the truncated Wigner approximation (TWA).
In this paper we do not include the fluctuations, because our purpose here
was to understand first the mean field behavior entailed by 
the experimental conditions.  

Our study has revealed the importance, in configurations similar to those in the experiment, of two features that are not present in previous, idealized studies of sonic analog black holes: (i) the condensate density increases towards the 
center of the atom trap and, (ii) as a result of this inhomogeneity, and the shape of
the trap potential, there is no single inertial frame in which the condensate is stationary. A key
consequence of the latter feature 
is that the white hole horizon, defined in the locally relevant sense explained below, 
is not at rest with respect to the black hole horizon. 

As a result of these features,
growth of a standing wave and emission of Hawking radiation both occur
purely hydrodynamically, in a way that appears to be similar to 
what was observed in the experiment. As described below,
we have identified the mechanisms producing these simulated phenomena,
and we find that the black hole laser mechanism plays no role. 
This raises doubts concerning whether lasing actually plays any role in the experiment.  
Even though our simulations do not take quantum fluctuations into account, the mean field behavior they reveal would persist in the presence of quantum fluctuations, and it already 
seems to account for the observed mean field behavior.
On the other hand, even if the conditions for lasing existed in the system, we may not have seen that instability because of the absence of the necessary seed fluctuations. 

There is an important feature of the experimental measurements that our simulations in this paper do not address, 
and that is the connected density-density correlation function. This is because our simulations are deterministic.
In order to understand the relation between this correlation
function and the mean field behavior, as well as to check
for lasing instability in the presence of quantum fluctuations, 
in another paper \cite{correlations} we have studied the system in the presence of both quantum fluctuations, and fluctuations in atom number from one run to the next. 
We find there that the correlation function is engendered by fluctuation-induced modulation of the deterministic standing wave. In fact, it results primarily from the varying number of atoms in the condensate.
When averaging over GP simulations with a 10-20\% variation in the number of atoms, and no quantum fluctuations, a correlation function similar to the measured one is produced. 
The addition of quantum fluctuations improves the agreement, but 
their effect is sub-dominant.
That is, we find that the observed correlation arises by modulation of the standing wave, and is unrelated to the intrinsic quantum correlations that would be present without the standing wave. In particular, we find that the quantum fluctuations do not trigger any instability.

Returning now to the results of the simulations reported here, let us summarize our findings.
As the potential step is swept through the condensate, 
a growing hydrodynamic standing wave (Fig.~\ref{fig:GP1}) arises between the horizons, 
which appears similar to the
one observed in the experiment. This wave is 
Bogoliubov-\v{C}erenkov radiation (BCR)~\cite{Carusotto},
generated at the WH~\cite{Coutant:2012zh,Busch:2014hla,Mayoral2011} [see Figs.~\ref{fig:exp}(d) and (e)]. 
It is also known as a zero-frequency undulation, and is reminiscent of a ship's bow wave.

Three independent lines of evidence all indicate the BCR nature of this standing wave.
First, the growth rate of the standing wave (Fig.~\ref{fig:GP_growth}) matches very closely the square of the background density, 
which changes as the step sweeps into denser parts of the BEC,
as would be expected from the BCR mechanism. (The time dependence of this 
growth also appears roughly compatible with that observed in the experiment.) Second, a spacetime plot of the magnitude of the deviation of the GP wave function from the background flow
[Fig.~\ref{fig:stimulated_HR}(b)] shows that the standing wave first arises at the WH, and then propagates to the BH [as illustrated in Fig.~\ref{fig:stimulated_HR}(a)].
And third, the standing wave has zero frequency in the WH frame. This is evident by inspection of
Fig.~\ref{fig:stimulated_HR}(b), which shows that the lines of constant phase are parallel to the WH horizon worldline.

Figures~\ref{fig:stimulated_HR}(a) and (b) also reveal that the WH horizon has a smaller velocity than the BH horizon. 
The BCR is therefore Doppler shifted to a nonzero frequency in the BH frame 
(which is the rest frame of the potential step), as can also be seen in the figure.
Although the relative velocity of the horizons is small, this Doppler shift is larger than 
might be expected, because the BCR has a very large wave vector. With its nonzero frequency in the BH frame, the BCR classically stimulates emission of 
Hawking radiation at the BH horizon, i.e., ``pair production" of an outgoing wave and partner radiation inside the horizon, as seen in Fig.~\ref{fig:stimulated_HR}(b).
We verified that the partner has the same frequency as the Doppler-shifted BCR, using a temporal windowed Fourier transform, Figs.~\ref{fig:stimulated_HR}(c) and (d).
This shows that only two frequencies are present: 
that of the background flow, and that of the BCR.
It may be possible to perform a similar analysis on the experimental data. 

We also explored trap parameters near those that roughly matched the experiment, 
seeking a regime that could yield a more distinct signal both for our analysis and 
in future experiments, We identified a slightly modified regime, in which all features are
qualitatively the same as those in the experimental regime, only much sharper [see Fig.~\ref{fig:M20}(b)]. 
In the modified regime we made a thorough spectral analysis of the condensate, using both
temporal and spatial windowed Fourier transforms [Figs.~\ref{fig:STFT_wt}(b) and (c)]. This enabled us to establish that,
despite the significant inhomogeneity in the system, 
the Bogoliubov--de Gennes dispersion relation gives a remarkably accurate prediction for the 
temporal and spatial spectral content of the BEC, with the only inputs being (1) the assumption
that the BCR has zero frequency in the WH frame, and (2) the local sound speed and flow velocity.
[We used spatial windowed Fourier transform of the density and of the GP wave function 
to identify the background flow and determine its density (for the sound speed) and 
wavevector (for the velocity).]
This detailed spectral analysis
gives us confidence that there is nothing going on beyond the mechanisms we have identified.
It also allows us to establish that the Hawking temperature prediction is consistent with the relative amplitude of the Hawking radiation and partner waves, insofar as would be expected. 
More generally, it reveals the utility of windowed Fourier transform in characterizing the 
local structure of an inhomogeneous BEC flow.

Finally, we conclude this introduction by mentioning that related studies having some overlap with ours have been reported in Refs.~\cite{Tettamanti, SR}. We comment on the relation between that work and our conclusions in Sec.~\ref{sec:comparison}.

\section{Methods}
\label{sec:method}

Reference~\cite{nphys3104} reported a step-sweeping experiment on a quasi-one-dimensional (quasi-1D) condensate. 
A detailed discussion of our approach to simulating this experiment is given in Appendix~\ref{app:sim_details}.
We describe a three-dimensional (3D) model of the experimental potential and condensate in Appendix~\ref{app:1dGPE}, while in Appendix~\ref{app:3dGPE} we 
discuss and evaluate the criteria for 
applicability of the reduction to a 1D model, and
compare with a simulation using the 3D GPE.
We find that the qualitative features in the 1D simulation are consistent with the results of 3D simulation. This indicates that the results of our detailed analysis of the 1D simulation should apply as well to the 3D system. 

The GP wavefunction, $\Psi(x,t)$, in our simulation involves a condensate component and the excitation modes generated during the sweep:
\begin{eqnarray}
\label{eq:psi0}
\Psi(x,t)=\Psi_{\rm bf}(x,t)+\sum_{j}\psi_{j}(x,t),
\end{eqnarray}
where $\Psi_{\rm bf}$ indicates the background condensate flow, and $\psi_{j}$ denotes its excitation mode satisfying the Bogoliubov--de Gennes (BdG) equation \cite{pethick}. In our simulation, there are three modes that we observed: the BCR mode, the HR mode, and the HR partner (labeled by $j=$ BCR, HR, and p, respectively). The role of each mode will be explained in the later sections. In regions where the flow is slowly varying, each component in Eq.~\ref{eq:psi0} behaves locally as a Wentzel-Kramers-Brillouin (WKB) plane wave with a characteristic frequency $\omega$ and wavevector $k$.  Here we introduce two techniques to resolve individual components and their spectral properties.

\subsection{Windowed Fourier transform}
\label{sec:WFT0}
A windowed Fourier transform (WFT) \cite{gomes99fourier} is a method that brings out the ``local" spectral elements of a function in the neighborhood of a given position or time.  It differs from the normal Fourier transform by including a Gaussian function centered at the position ($x$) or time ($t$) of interest. The spatial WFT $F(k,x)$ of a function $f(x)$ is defined as:
\begin{eqnarray}
\label{eq:FTkx0}
F(k,x)= \int^{\infty}_{-\infty} dy\, f(y)w(y-x;D)e^{-iky},
\end{eqnarray}
where $w(y-x;D)=\exp (-(y-x)^2/D^2)/\left(\sqrt{\pi}D\right)$ is a Gaussian window function of width $D$. With the filtering of the window, the transformed function $F(k,x)$ constitutes a local Fourier transform of $f(x)$, capturing features that vary on length scales much smaller than $D$.  For instance, given a function, $f(x)=f_{q}(x) \exp (iqx)$, with wavevector $q$ and slowly varying amplitude $f_{q}(x)$, the transformed function is $F(k,x)\approx f_{q}(x) \exp (-(k-q)^2(D/2)^2)$ : a Gaussian in $k$-space, centered at $k=q$ with width $2/D$, and the peak height is the local amplitude, $f_{q}(x)$. 

The WFT is able to resolve locally (at a given $x$ or $t$) the Fourier components in Eq.~\ref{eq:psi0} as peaks in the resulting wavevector (or frequency) spectrum, in which peak position and height indicate the wavevector and amplitude of each component. Specifically, for a background condensate flow, $\Psi_{\rm bf}\sim |\Psi_{\rm bf}(x)|e^{ik_{\rm bf}(x)x}$, its spatial WFT exhibits the local wavevector $k_{\rm bf}(x)$ for each $x$, which determines the local flow velocity in the laboratory frame, $\hbar k_{\rm bf}(x)/m$. This is shown as the main streak in Fig.~\ref{fig:exp}(b).  Similarly, the spatial WFT of the density, $n(x)=|\Psi(x)|^2$, separates the background condensate density $n_{\rm bf}(x)$  ($k=0$) from the superimposed spatial oscillations (with nonzero $k$). An example is shown in  Fig.~\ref{fig:GP_growth}(b), which is the spatial WFT of density profile in Fig.~\ref{fig:GP1}(g), evaluated at the center of the oscillatory region.  The spectrum has a central peak $n_{\rm bf}$ as the background density, and two side peaks $n_k$, indicating the oscillatory component. 

\subsection{Moving average of the GP wavefunction}
\label{sec:smoothing0}

To separate fast-oscillating components in Eq.~\ref{eq:psi0} from the slowly-varying parts, we implement a smoothing procedure on the GP wavefunction. The procedure is equivalent to calculating the moving average of a discrete data set, which smooths out short-range fluctuations. Here the moving average of wavefunction $\Psi(x)$ is defined as
\begin{eqnarray}
\label{eq:smoothing0}
\bar{\Psi}(x)=\frac{1}{2D_s} \int^{x+D_s}_{x-D_s} dy\, \Psi(y),
\end{eqnarray}
where the integral serves as a square window of width $2D_s$ centered at $x$, over which $\Psi(x)$ is being averaged. For components in $\Psi(x)$ with wavelength shorter than $D_s$ (i.e., $D_s>\pi/k$), the integral would give rise to an average of zero, leaving those that are slowly varying in space (i.e., $\pi/k > D_s$) in $\bar{\Psi}(x)$, and the difference $\delta\Psi \equiv \Psi -\bar\Psi$ characterizes the part of 
$\Psi$ composed roughly of wavevectors $k\gtrsim \pi/D_s$.  Later in Sec.~\ref{sec:BCRStim}, we use the above procedure at each time and exhibit a spacetime diagram of $|\delta\Psi(x,t)|$, in which the slowly-varying part of the background flow ($\Psi_{\rm bf}$ with $k\sim 0$) is removed to bring out $\psi_{\rm HR}$.

\section{Analysis of the simulated experiment}
\label{sec:experiments}
\subsection{Formation of the BH-WH cavity }
In the experiment of~\cite{nphys3104}  a BH-WH cavity is established in a quasi-one-dimensional, laboratory BEC held by a confining potential, $U(x)$, as shown in Fig.\,\ref{fig:exp}(a). By sweeping a potential step of depth $U_{s}$ at uniform speed $v_{s}$ across the BEC, BH and WH are established ($U_s$ is on the order of $10^{-9}$ K, and $v_s$ is 0.21 mm/s).  Atoms are accelerated in the direction opposite to the step motion due to the precipitous drop in the potential.  This creates a supersonic flow behind the step and forms a BH at the step edge, $x_{\mathrm{BH}}$. The accelerated atoms gradually slow as they recede from the step, due to the rising potential. This causes the flow to become subsonic at a critical distance $L$ behind the step, forming a WH, $x_{\mathrm{WH}}$. Not far beyond $x_{\rm WH}$ the flow velocity in the laboratory frame drops to zero, roughly where $U(x)-U_{s}=U(x_{\mathrm{BH}})$ (this implies that $L$ increases slightly as $x_{\mathrm{BH}}$ moves toward the center of the trap). This procedure produces the flow structure shown in Fig.~\ref{fig:exp}(c).

To determine the flow structure, we implement the spatial WFT described in Sec.~\ref{sec:WFT0}. Figure~\ref{fig:exp}(b) is a local wavevector spectrum $|\Psi(k,x)|^2$ with $D=5$ $\mu$m, defined in the laboratory frame at a moment during the sweep [Fig.~\ref{fig:GP1}(e)]. There is a dominant streak, indicating the background condensate flow, $\Psi_{\rm bf}$, for which the peak position at each $x$ defines the background wavevector, $k_{\rm bf}(x)$. The regions with zero wavevector, $k_{\rm bf}\sim 0$, correspond to the non-accelerated, subsonic BEC; the region behind the step with $k_{\rm bf}\sim -1.4$ $\mu$m$^{-1}$ corresponds to the accelerated, supersonic flow. The blue (dark gray) curve in Fig.\,\ref{fig:exp}(c) is minus the flow velocity in the rest frame of the step, $v(x)=v_{\rm bf}(x)-v_{\rm s}$, where $v_{\rm bf}(x)=\hbar k_{\rm bf}(x)/m$ is the background flow velocity in the laboratory frame. The green (light gray) curve is the local speed of sound $c(x)=\sqrt{g{n_{\rm bf}}(x)/m}$, where $g$ is a 
coupling constant defined in Appendix \ref{app:1dGPE}, and 
${n_{\rm bf}}(x)$ is the local density of the background flow, which we identify here using
a WFT of the density $n(x)$ (see Appendix \ref{app:FT_vc} for details). 

\subsubsection{Locating the black and white hole horizons}
\label{sec:Locating}
The black hole horizon is defined as the location where a right moving phonon is at rest in the step frame. This corresponds to the right intersection of $c(x)$ and $-v(x)$ in Fig.~\ref{fig:exp}(c). The step frame is distinguished as the one in which the system is closest to being stationary near the step. 
In particular, $c(x)$ and $v(x)$ are nearly steady where the trapped BEC spills over the moving step.

The definition of the WH horizon is not as simple, because the step frame is not a global stationary frame of the system, due to the spatial variation in the trap potential and background condensate density. 
Instead, what is 
dynamically important is the location of the transition from supersonic to subsonic in the frame in which conditions are {\it locally} stationary. In particular, this is the locus of \v{C}erenkov radiation, which arises from the accessibility of negative energy modes in a frame in which energy is conserved, i.e., in which conditions are stationary.

At early times in the sweep of the step, the WH horizon so defined is located where the BEC density, and therefore the sound speed, is significantly smaller. It therefore starts out moving much more slowly than the step. As the sweep progresses it accelerates smoothly, until it reaches a uniform velocity slightly less than that of the step. It is then approximately located at the left intersection of $c(x,t)$ and $-v(x,t)$, and the distance $L$  between the two horizons grows slowly and uniformly in time. This behavior 
can be seen in Figs.~\ref{fig:stimulated_HR}(b) and \ref{fig:M20}(b).

\subsection{Cavity standing wave}

\begin{figure}[htb]
\includegraphics[width=3.5in]{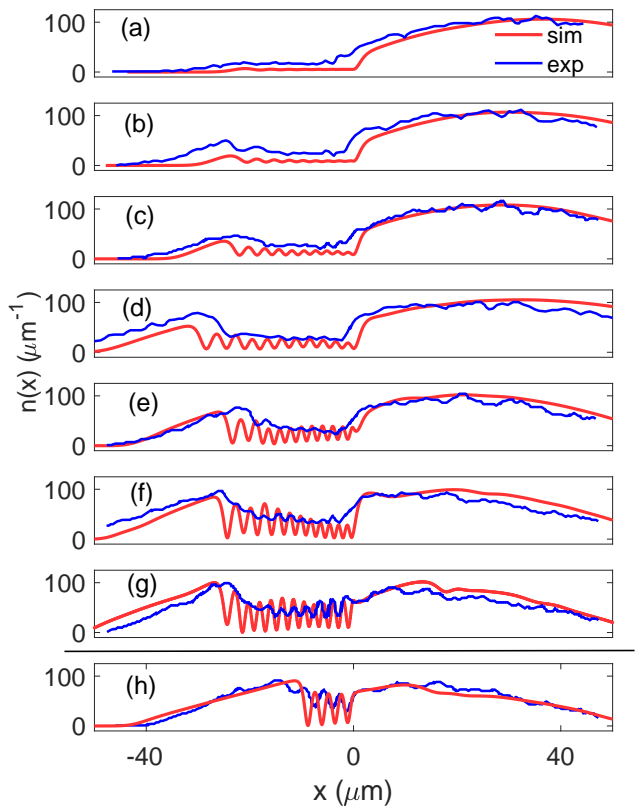}
\caption{\label{fig:GP1}  
(a-g) Density vs. time of a swept BEC at 20 ms intervals with step $U_s/k=\text{ 6 nK}$,
scaled by a common factor to match experiment, and viewed in the moving frame where $x=0$ defines the step edge; panel (h): $U_s/k =\text{ 3 nK}$ at 120 ms. Blue (dark gray): experiment \protect\cite{nphys3104}; red (gray): present simulation. }
\end{figure}

Figure\ \ref{fig:GP1} shows comparisons of the simulated density profile with experiment. Figures (a)-(g) show the BEC density for $U_s = k \times 6$ nK after the launch of a sweep at 20 ms intervals, where $k$ is the Boltzmann constant. Figure (h) corresponds to the density profile at $t=$ 120 ms for $U_s = k \times 3$ nK. The coordinate origin in each panel has been displaced to coincide with $x_{\rm BH}$.

The density exhibits a standing-wave pattern behind the step, with amplitude growing in time. 
Considering that the experimental observations involve an average over any quantities that fluctuate from one run to another, 
the GP simulation qualitatively matches the overall evolution 
seen in the experiment. 
In particular, the growth, wavelength, and phase of the wave pattern are similar to each other.

\subsection{\v{C}erenkov mechanism}
\label{sec:mechanism}

In the following, we present multiple lines of evidence showing 
that the standing wave 
results from the Bogoliubov-\v{Cerenkov} radiation (BCR) effect,
in a process closely
analogous to the flow past an obstacle studied in \cite{Carusotto}. 
This evidence
is based on the wavevector and frequency 
spectra of the standing wave, and the growth rate of the standing wave, which
we will show is due to the increasing BEC density. We also establish that the Hawking radiation is emitted by this system and that the partner mode slightly modulates the standing wave.

As illustrated in Fig.\ref{fig:exp}(d), an obstacle in a stationary supersonic flow produces an upstream, Bogoliubov-\v{Cerenkov} standing wave \cite{Leboeuf2001,Carusotto}, analogous to a bow wave on water \cite{Carusotto:2012fy}. It was observed in Ref.~\cite{Mayoral2011} that such a standing wave is generated at a WH, triggered by an incident wavepacket on the stationary flow, and saturating at an amplitude determined by nonlinear effects. 
A similar standing wave, generated by inhomogeneity at a WH horizon, can be seen in Fig. 4 of Ref.~\cite{PhysRevA.91.053603}.
In our case, the subsonic component to the left of the WH serves as an obstacle in the supersonic flow, generating a Bogoliubov-\v{Cerenkov} wave.

\subsubsection{Wavevector spectrum}
\label{sec:kspectrum}

In the WFT spectrum in Fig.~\ref{fig:exp}(b), we observe in the cavity region an excitation mode at $k\sim1.4$ $\mu$m$^{-1}$ coming from the WH, which is roughly the reflection of the supersonic flow $\Psi_{\rm bf}$ with  $k\sim-1.4$ $\mu$m$^{-1}$. The interference of the two results in the standing wave in the density profile shown in Fig.~\ref{fig:GP1}(e), which has a wavevector with twice the above value,  $k\sim3$ $\mu$m$^{-1}$. 

The relation  $k_{\rm BCR} \sim-k_{\rm bf}$
is expected from energy conservation: 
the flow structure is approximately time-independent in the rest frame of the WH horizon, so we expect that the BCR production process should conserve energy. One way to view it is that incoming atoms reflect from the flow transition at the horizon. The velocity of the WH horizon is quite low compared to the flow velocity upstream, so to a good approximation energy conservation in the WH frame implies that the laboratory frame wavevector should simply reverse sign. A more precise account of this given in Sec.~\ref{sec:M2} below, using a linearized mode analysis.

Another consequence of the approximate local time independence in the WH frame is that the frequency of the standing wave generated there should have zero frequency in that frame. Indeed it does, but we postpone the demonstration of that to later in this section.
 
\subsubsection{Growth of the standing wave}
\label{comparison}

\begin{figure}[tb]
\includegraphics[width=3.in]
{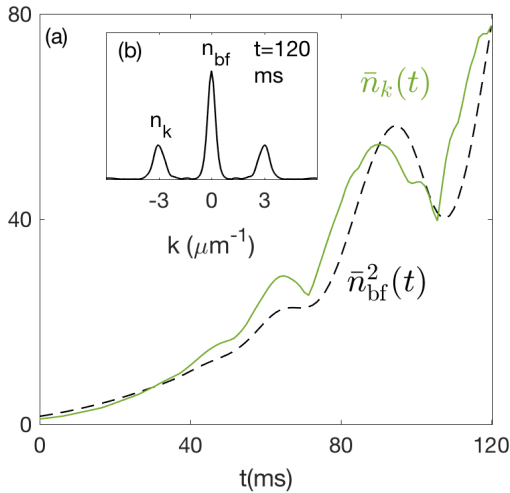}
\caption{\label{fig:GP_growth}  
(a) Simulated growth of the standing-wave pattern in the supersonic region for $U_{s}/k =$  6 nK. Solid green: normalized standing-wave amplitude $\bar{n}_k(t)$, $\bar{n}_{k}(t)=n_{k}(t)/n_{k}(0)$, for which $\ln[\bar{n}_{k}(120)]\sim 4.4$. 
Dashed black: the square of background density, $\bar{n}_{\rm bf}(t)$, scaled to match the final standing-wave amplitude, $\bar{n}_{\rm bf}^2(t)=n_{\rm bf}^2(t)[\bar{n}_{k}(120)/n_{\rm bf}^2(120)]$. The growth of $n_{\rm bf}$ and $n_{k}$ is determined from a spatial WFT of $n(x)$ at $x=-12$ $\mu$m with window width $D=6.5$ $\mu$m. Inset (b)
shows the windowed wavevector spectrum at $\text{t=120 ms}$.}
\end{figure}

Reference~\cite{nphys3104} reported exponential growth of the oscillatory density 
pattern in the BH-WH cavity, and suggested that it results from the black hole laser effect.
Our simulations exhibit similar growth, but lead us to attribute it to a 
different mechanism.
Figure~\ref{fig:GP_growth} 
displays 
the growth of the background flow density $n_{\rm bf}$ and of a standing wave, $n_k$, defined by the peaks of the WFT 
of the density at $x_{\rm BH}-12 \mu$m, as shown in the inset. Note that this spectrum is different from the one in Fig.~\ref{fig:exp}(b), which is the squared modulus of the spatial WFT of the wavefunction. Over 120 ms the standing wave density grows by $\sim \exp(4.4)$. 
Figure~\ref{fig:GP_growth} shows that $n_k$ grows in proportion to $n_{\rm bf}^2$. The two oscillation features superimposed on the growth curve coincide with the variations seen in Fig.~\ref{fig:stimulated_HR}(b), and will be discussed in Sec.~\ref{sec:comparison}.

To understand this quadratic relationship between the 
standing wave amplitude and the background flow density, we begin by noting that
the step moves toward the region of higher BEC density [see Fig.~\ref{fig:exp}(a)], so the background density $n_{\rm bf}$ also grows in the cavity.
Now the saturated amplitude ($n_k$) 
of a BCR standing wave
should be proportional to both the strength of the obstacle ($V_{\rm ob}$), and the density of the background flow ($n_{\rm bf}$) \cite{Carusotto, pethick} [see Fig.\ref{fig:exp}(e)]. The ``obstacle" in the present case has a strength proportional to the BEC density to the left of the WH, which grows similarly to that on the right, so it follows that the saturated wave amplitude $n_k$ should grow as $n_{\rm bf}^2$.

The very close agreement with this scaling relation displayed in Fig.\ref{fig:GP_growth}(a), 
gives further
compelling evidence that the standing wave observed inside the supersonic cavity is in fact BCR, and indicates that its growth results from the increase of background density, rather than from 
a black hole laser instability.
Moreover, effects due to classical or quantum fluctuations, not included in our simulation, could not remove this robust, large BCR wave, but rather would have to appear in addition to it. 

The absence of fluctuations in our simulation here implies that we are
unable to capture the behavior of the 
density-density correlation function.
That correlation function was measured in the experiment \cite{nphys3104}, and displays a checkerboard pattern with periodicity
very close to that of the standing wave. The growth of the checkerboard pattern was 
quantified in  \cite{nphys3104} via the Fourier power spectrum of the correlation, and found to grow by a factor $\sim\exp(3.3)$. 
In \cite{correlations} we have shown, by introducing quantum and atom-number fluctuations into our simulations,
that this checkerboard pattern results directly from the presence of the underlying BCR standing wave, modulated by the fluctuations.

\subsubsection{Spacetime portrait}
\label{sec:BCRStim}

In this subsection we present a spacetime portrait for the evolving BEC. This portrait illustrates by 
visual inspection  that the standing wave is generated from the WH, 
and has zero frequency in the WH reference frame. Its frequency in the step frame is nonzero, due to a 
Doppler shift arising because the WH recedes from the BH as the system evolves. 
The spacetime portrait also reveals a signal of Hawking radiation, which is stimulated by the BCR at the BH.
We further verify this mechanism quantitatively through a windowed frequency spectrum evaluated inside the cavity,
which reveals that the only frequencies present are those of the background condensate and the BCR.

\begin{figure*}[htb]
\centering
\includegraphics[width=7in]{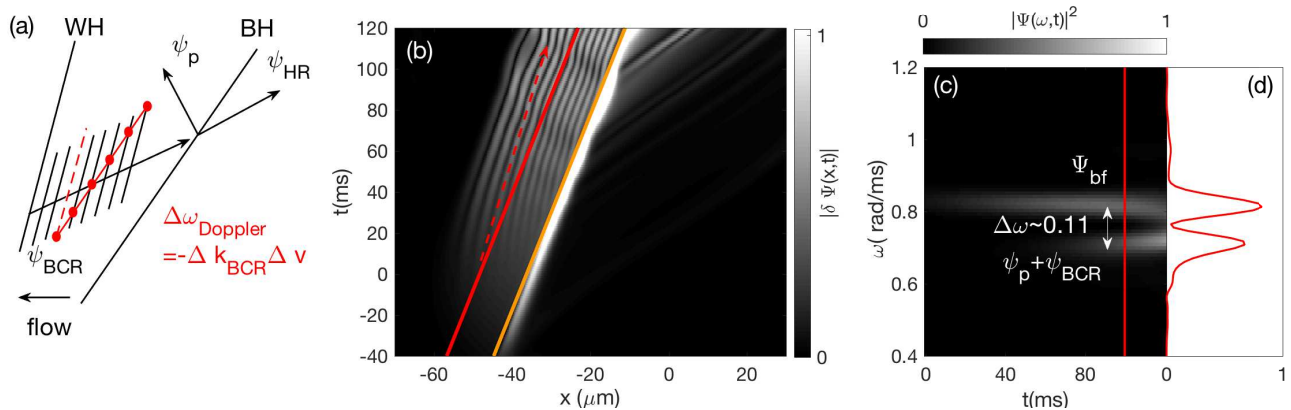}
\caption{\label{fig:stimulated_HR} 
Pair production, stimulated by a Doppler-shifted BCR mode. (a) Spacetime portrait (viewed in the laboratory frame) of HR ($\psi_{HR}$) and its partner ($\psi_{\rm p}$) created at the BH by right-propagating BCR ($\psi_{\rm BCR}$) generated at the WH as a standing wave. The BCR standing wave has zero frequency in the WH frame, but since the WH is receding from the BH by velocity difference $\Delta v$, the BCR in the BH frame has a nonzero frequency, $\Delta\omega= -\Delta k_{\rm BCR}\Delta v $, as can be seen by the phase change of the standing wave along the solid red (left) line. The Hawking pair is stimulated by the BCR at this nonzero frequency at the BH. (b) Time evolution of $|\delta \Psi(x,t)|$ in the experimental regime~\protect\cite{nphys3104}, multiplied by 10 for $x > x_{\mathrm{BH}}$. As in (a), the dashed red line is parallel to the WH worldline, and the solid red (left) line is parallel to the BH worldline, indicated by the diagonal orange (right) line. (c) Windowed frequency spectrum evaluated along the solid red (left) line in (b); (d) is the cut-through of the spectrum at  $t= 100$ ms.}
\end{figure*}

The spacetime portrait for the simulated experiment, Fig.~\ref{fig:stimulated_HR}(b), displays $|\delta\Psi|=|\Psi-\bar\Psi|$.
To resolve the HR outside the cavity, we subtract the moving average $\bar\Psi$.
using the procedure described in Sec.~\ref{sec:smoothing0}. 
$\bar\Psi$ approximates the dominant, slowly-varying background [with $k\sim0$ as in Fig.~\ref{fig:exp}(b)]. We choose the smoothing window $D_s=5.4\ \mu$m, such that it is large enough so that $\bar\Psi\approx0$ between the horizons, yet small enough to capture the slow variations of the background outside the horizon.

The portrait displays an interference pattern between the background supersonic flow and excited modes of $\delta\Psi$. The evolution of the BH is indicated by the diagonal orange (right) line. To clearly display HR upstream of the horizon, we have multiplied $|\delta\Psi|$ there by a factor of 10. At the beginning of the evolution, as the condensate spills over the step, a left-moving flow develops, indicated by the growing light gray area. When this flow reaches the WH, at $t\approx 10$ ms,  
a standing wave (BCR) is generated. 
In Fig.~\ref{fig:stimulated_HR}(b), it is clear by inspection of the dashed red line and solid red (left) line [which is parallel to diagonal orange (right) line] that the standing wave has zero frequency in the WH rest frame, but nonzero frequency in the BH rest frame. Since its frequency is nonzero in the BH frame, the BCR can stimulate production of Hawking pairs at the BH horizon. (Zero frequency Hawking pairs do not exist.)

The stimulated HR is seen in the spacetime portrait
Fig.~\ref{fig:stimulated_HR}(b). The BCR first reaches the BH at $t\approx 20$ ms, stimulating emission of HR. 
Hawking radiation first appears at around 25 ms, but is not visible on the 
grayscale plot until
around 40 ms.
The left--moving partner radiation (p-mode) resulting from the 
``pair creation" forms a ``V"-shape with the HR, and makes
an interference pattern with the BCR that can first be seen around 
$t\approx 40$ ms.
(Were there no mode present to stimulate the pair creation,
it would nevertheless occur spontaneously, as in the Hawking effect for an astrophysical black hole.) 

Figure~\ref{fig:stimulated_HR}(a) is a schematic illustration of the 
mechanism just described. This is viewed in the laboratory frame, where the BH moves at velocity $v_s$ and the WH with a slightly smaller velocity, $v_s-\Delta v$, as indicated by the dashed red line.
As seen by inspection of Fig.~\ref{fig:stimulated_HR}(b),
the BCR  (i.e.,\ the standing wave) has zero phase velocity with respect to the WH,  corresponding to zero frequency in the WH frame.
Since the WH velocity is less than the BH velocity [as shown in Figs.\ref{fig:stimulated_HR}(a) and (b)], this gives rise to a nonzero frequency in the BH frame. 
Note that, although the relative velocity of the BH and WH is rather small, 
the BCR wavelength is rather short, so that
the BCR frequency in the BH frame is not small.  
As the BCR mode ($\psi_{\rm BCR}$) propagates to the BH, stimulates the emission of HR ($\psi_{\rm HR}$) and its partner ($\psi_{\rm p}$)  at the latter frequency, with the associated wavevectors determined by the Bogoliubov-de Gennes (BdG) spectrum.

\subsubsection{Frequency spectrum}

In Fig.~\ref{fig:stimulated_HR}(c) and (d), we show the windowed frequency spectrum in the experimental regime in the supersonic region. The frequency is computed in the rest frame of the moving step, along the solid red (left) line in Fig.~\ref{fig:stimulated_HR}(b). The long streak, which starts from the beginning of the evolution, corresponds to the background flow, $\Psi_{\rm bf}$. The short streak, which is separated from the long streak by $\Delta\omega\sim 0.11(3)$ rad/ms, corresponds to the BCR and the p-mode. $\Delta\omega$ is nonzero because of the Doppler shift between the WH and BH frames.  

The Doppler effect due to the recession of the WH can be estimated using the velocity difference between the two horizons, $\Delta v\sim0.03$ mm/s. The shifted frequency is the product of $\Delta v$ and the BCR wavevector, $\Delta k_{\rm BCR}$,
\begin{equation}\label{eq:shift}
\Delta \omega=-\Delta k_{\rm BCR}\Delta v=- 0.09 \mbox{ rad/ms}.
\end{equation}
We find that the WFT frequency agrees with the prediction in Eq.~\ref{eq:shift} to within the uncertainty. 

The quantitative agreement with zero WH frequency shifted to the BH frame establishes that the mechanism illustrated in Fig.~\ref{fig:stimulated_HR}(a) is operative. In particular, the Doppler shifting of frequency between the two horizons plays an essential role in the process, and the partner mode of the Hawking radiation has the same frequency as that of the BCR.
The fact that the partner waves match this frequency shows that they are stimulated by the BCR, rather than being self amplifying.

\section{Enhanced parameter regime}
\label{sec:M2}

In addition to simulating the system using parameters close to those of the experiment, we have explored a different parameter regime, in which the phenomena observed in the experiment, in particular the Hawking radiation, are more sharply displayed. This was helpful in developing an understanding of the behavior of the system, and it may prove useful for optimizing the Hawking radiation signal in future experiments. 

In the experimental regime, the signal of HR is too weak to be directly seen in the density profile in Fig.~\ref{fig:GP1}. With the help of the spacetime portrait in Fig.~\ref{fig:stimulated_HR}(b), one observes emission from the BH that resembles HR, but with irregularity. This irregularity may be due to the long wavelength of the p-mode, $\lambda_{\rm p}$, relative to the cavity size, $L$. Since $\lambda_{\rm p}\sim L$, the p-mode does not behave as a WKB mode on a slowly-varying background. This, in concert with the time dependence of the cavity size,
may lead to the irregular emission of the HR mode.

\begin{figure}[htb]
\includegraphics[width=3.1in]{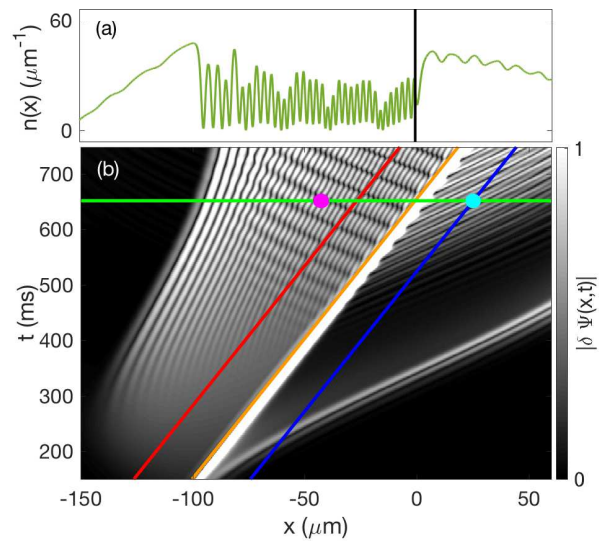}
\caption{\label{fig:M20}
Stimulated pair production in the enhanced regime, M2. Panel (a): density $n(x)$ at $t = 650$ ms, along the horizontal green line in (b); panel (b): spacetime portrait. The diagonal red (left) and blue (right) lines indicate the paths on 
which the windowed frequency spectra of Fig.~\ref{fig:STFT_wt}(b) are 
calculated. The wavevector spectrum along the horizontal green line is shown in  
Fig.~\ref{fig:STFT_wt}(c). The magenta (left) dot and cyan (right) dot correspond to a correlated Hawking pair, for which the thermal prediction is being tested.}
\end{figure}

In the enhanced parameter regime, we lower the $\lambda_{\rm p}/L$ ratio by modifying the parameters of the trapping potential (axial trap frequency $\omega_x$) and the step potential ($U_s$ , $v_s$). 
Figure~\ref{fig:M20}(a) shows the density profile in one such modified regime (case M2), from which sharper signals of HR and p-mode have been observed, with suppressed $\lambda_{\rm p}/L$ ratio.  In this case, the BEC is twice as long as in the experiment of \cite{nphys3104}, the step size is halved relative to the 6 nK step, and the step speed is about the same. The details of the investigation of parameter regimes are summarized in Appendix~\ref{app:parameter}. 

The spacetime portrait of the modified regime is shown in Fig.~\ref{fig:M20}(b). The BCR-stimulated pair production mechanism illustrated in Fig.~\ref{fig:stimulated_HR}(a) can be seen very clearly, with more distinct features than in the experimental regime [Fig.~\ref{fig:stimulated_HR}(b)]: (i) the BCR, with phase parallel to the WH, which grows substantially prior to the pair creation, (ii) the ``V"-shaped HR pair, stimulated by the BCR. Note that the frequency of $|\delta\Psi|$ appears doubled outside the BH compared to that inside. This is because $\delta\Psi$ contains very little background flow component with which to interfere outside the BH, so the visible interference is between the positive and negative relative frequency parts of the HR. 

Furthermore, since the HR and the p-mode have enhanced signals and regular wavelengths here, their spectral properties can be captured by WFTs. In the following, we analyze the properties of the modes based on the Bogoliubov-de Gennes theory.  

\begin{figure*}[!ht]
\centering
\includegraphics[width=7.3in]{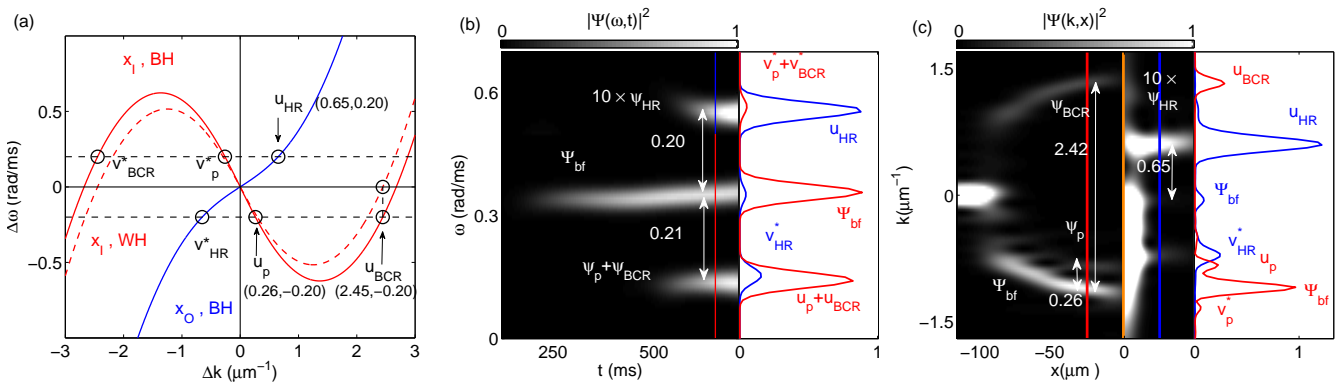}
\caption{\label{fig:STFT_wt} 
Dispersion relations and WFTs of $\Psi(x,t)$ [$\delta\Psi(x,t)$ for $x>x_{\rm BH}$] along the lines indicated in Fig.~\ref{fig:M20}(b). (a) Dispersion relations at $x_{\rm I}$ in the WH (dashed red) and BH (solid red, or gray) reference frames, and at $x_{\rm O}$ in the BH reference frame (solid blue, or dark gray), evaluated at $t=$ 650 ms. 
$\Delta \omega$ and $\Delta k$ are the frequency and wavevector relative to those of the background flow. (b) Frequency spectrum along the diagonal red (left) ($x_{\rm I}$) and blue (right) lines ($x_{\rm O}$);  (c) wavevector spectrum along the
horizontal green line ($t=$ 650 ms).}
\end{figure*}

\subsection{BdG mode analysis}
\label{sec:BdG} 
The BdG theory of linearized modes \cite{pethick} (Appendix~\ref{app:BdG})
can be used to predict the temporal and spatial WFT spectra of the BEC, starting from only one input assumption: that the standing wave has zero frequency in the WH frame. This will further verify the mechanism we have proposed for the excitations of the BEC. In addition, it will demonstrate the remarkable accuracy of BdG analysis when combined with WFT in an inhomogeneous setting.

To make contact with the notion of 
linearized, BdG modes and their dispersion relation, 
we locally factor the full GP wavefunction $\Psi(x,t)$ into a homogeneous background $\Psi_{\rm bf}$
and the deviation $\psi$, so that the deviation is locally a superposition of harmonic modes of the form 
\begin{eqnarray}
\label{eq:BdG1}
\psi_j &=& \left( u_{j} e^{-i\D\omega_{j} t+i \D k_{j}x} +v_{j}^{*} e^{+i\D\omega_{j} t-i \D k_{j}x}\right)\nonumber\\
&& ~~~\times e^{-i\omega_{\rm bf} t+ik_{\rm bf}x},
\end{eqnarray} 
where $j=$p, HR, BCR.
Each BdG mode is composed of two components, with opposite frequency and wavevector, 
$\pm(\D\omega_j,\D k_j)$, relative to those of the background flow, 
$(\omega_{\rm bf},k_{\rm bf})$. The BdG dispersion relation is given by
\begin{eqnarray}
\label{eq:dispersion}
\Delta\omega &=&\sqrt{c^2\Delta k^2+ (\hbar \Delta k^2/2m)^2}+v_{\rm bf, o}\Delta k,
\end{eqnarray}
where $v_{\rm bf,o}$ is the velocity of the condensate with respect to the ``observer" frame
in which the frequency is defined.
The square root term gives the frequency in the comoving frame of the condensate, $\Delta\omega_{\rm cm}$. 
The first term in the square root describes long wavelength sound modes, while the second term corresponds to the kinetic energy of the atoms, and dominates at large wavevectors. 
The amplitudes of two components of $\psi_j$ are given by
\begin{eqnarray}
\label{eq:uv}
(u_j,v_j)=\frac{1}{2\pi}\sqrt{\left|\frac{d\Delta k}{d\Delta\omega}\right|} \left(\frac{1}{\sqrt{1-D^2}},\frac{D}{\sqrt{1-D^2}} \right),
\end{eqnarray}
where $D=[\hbar\Delta\omega_{\rm cm}- \hbar^2 \Delta k^2/2m - mc^2]/mc^2$ \cite{parentani2010}. Note that $D$ goes to zero for $\Delta k\gg mc/\hbar=1/\sqrt{2}\xi$, where $\xi$ is the healing length.

In the enhanced regime, the dispersion relations evaluated inside and outside the BH at $t=650$ ms are shown in Fig.~\ref{fig:STFT_wt}(a). The red (gray) and blue (dark gray) solid curves indicate the dispersion relation in the BH frame ($v_{\rm bf,o}=v_{\rm bf,BH}$), at $x_{\rm I}=x_{\rm BH}-$26 $\mu$m and $x_{\rm O}=x_{\rm BH}+$26 $\mu$m, respectively. These points correspond to the intersections of the diagonal red (left) and blue (right) lines with the horizontal green line in Fig.~\ref{fig:M20}(b).
The dashed red curve also indicates the dispersion relation at $x_{\rm I}$, but referred to 
the WH frame ($v_{\rm bf,o}=v_{\rm bf,WH}$). We use the numerically measured values of the local flow velocity and sound speed, determined from the background flow $\Psi_{\rm bf}$, which can be identified by a spatial WFT (despite the appearance of additional excitations). The WH velocity is approximated by the speed of the left edge of $|\delta\Psi(x,t)|$ [see Fig.~\ref{fig:M20}(b) and Appendix~\ref{app:dispersion}], while the BH velocity is that of the step. 

The BCR has zero frequency in the WH frame, so the BCR wavevector should satisfy $\Delta\omega(\Delta k_{\rm BCR})=0$ in that frame. 
This is indicated graphically by the intersection of the dashed red dispersion curve in 
Fig.~\ref{fig:STFT_wt}(a) with the $\Delta k$ axis, 
which yields $\Delta k_{\rm BCR}\sim 2.5\, \mu{\rm m}^{-1}$. 
  
Due to the recession of WH relative to the BH, the frequency of BCR in the BH frame corresponds to $\Delta \omega=- \Delta k_{\rm BCR}\Delta v=- 0.2 \mbox{rad/ms}$, where $\Delta v$ is the BH velocity relative to the WH. This frequency is indicated by the lower dashed horizontal line, which intersects the solid red (gray) curve at the vertical line, $ \Delta k_{\rm BCR}$. (Note that similar reasoning can be applied for the upper dashed horizon line, which intersects the component, $v^*_{\rm BCR}$, at the opposite frequency and wavevector.)
If the HR and partner modes are indeed stimulated by the 
BCR, they should share the same frequency with the BCR in the BH frame, so their wavevectors should lie at the intersections of the shifted BCR frequency (dashed black) lines with the solid blue (dark gray) and solid red (gray) dispersion curves, respectively. 

As the BCR mode propagates toward the BH horizon, the dispersion curve lifts upwards due to the change of flow velocity and sound speed, and the wavevector ``redshifts", until the mode coincides with the local minimum of the dispersion relation. At that stage the WKB description breaks down, and the mode converts to a superposition of other modes that share the same frequency. 
These are the Hawking radiation and partner modes.
The modes are labeled by ``$u$" or ``$v^*$", according to the corresponding component of the BdG mode \eqref{eq:BdG1}. 
Modes whose $u$-component has negative (positive) relative frequency in the step frame have negative (positive) energy relative to the condensate \cite{PhysRevA.80.043601}. 
The BCR and partner modes thus have negative energy, while the Hawking mode has positive energy.

\subsection{Spectral comparison with BdG prediction}
\label{sec:spectralHR}

To capture the spectral properties of the modes observed in Fig.~\ref{fig:M20}(b), and compare with the prediction in Fig.~\ref{fig:STFT_wt}(a), we apply the spatial and temporal WFTs on $\Psi(x,t)$ and $\delta\Psi(x,t)$. On the left-hand side of the BH ($x<x_{\rm BH}$), we calculate the WFTs of $\Psi(x,t)$; on the right-hand side of the BH ($x>x_{\rm BH}$), we take $\delta\Psi(x,t)$ and multiply it by 10 to subtract the background and bring out the HR. The left panel of Fig.~\ref{fig:STFT_wt}(b) shows the windowed frequency spectra
of $\Psi(x_{\rm I}(t),t)$ ($\omega=\text{0--0.5 rad/ms}$) and $\delta\Psi(x_{\rm O}(t),t)$ ($\omega=\text{0.5--0.7 rad/ms}$), in the BH frame, 
along the diagonal red (left) and diagonal blue (right) lines in Fig.~\ref{fig:M20}(b) with Gaussian width $T=\text{55 ms}$. The streak in the center corresponds to the background flow $\Psi_{\rm bf}$, 
and indicates the frequency $\omega_{\rm bf}\sim \text{0.36 rad/ms}$. 
The two other streaks located symmetrically about the center correspond to HR 
($\omega\sim \text{0.56 rad/ms}$), 
and the BCR and the p-mode ($\omega\sim \text{0.15 rad/ms}$). The full frequency spectra at $t=$ 650 ms for $x_{\rm I}$(diagonal red line, left) and $x_{\rm O}$(blue, right) are shown on the right panel. 

The left panel of Fig.~\ref{fig:STFT_wt}(c) shows the windowed wavevector spectrum as a function of position, in the laboratory frame. It is defined by WFTs of $\Psi(x,t_0)$ ($x<x_{\rm BH}$) and $\delta\Psi(x,t_0)$ ($x>x_{\rm BH}$) at $t_0=650$ ms, along the horizontal green line in Fig.~\ref{fig:M20}(b), with width $D=$ 21 $\mu\text{m}$ for $\Psi$, and 12 $\mu\text{m}$ for $\delta\Psi$. 
The background flow spectrum between the horizons is centered on a large negative wavevector at each $x$, and extends from the BH to the WH. 
As in the experimental regime [Fig.~\ref{fig:exp}(b)], the BCR spectrum is roughly the reflection of the background flow, $\Psi_{\rm bf}$.
This feature was explained qualitatively in Sec.~\ref{sec:kspectrum}. Here we can explain it quantitatively, using the dispersion relation \eqref{eq:dispersion}. As can be seen in Fig.~\ref{fig:STFT_wt}(a), the point at $\Delta\omega=0$ 
in the WH frame
is close to the single-particle regime (i.e., $\Delta\omega_{\rm cm}\sim \hbar \Delta k^2/2m$), so that $\Delta k_{\rm BCR}\sim -2mv_{\rm bf, WH}/\hbar$. The flow
velocity $v_{\rm bf, WH}$ relative to the WH is approximately the same as the velocity in the laboratory frame, which is $\hbar k_{\rm bf}/m$. Therefore 
$\Delta k_{\rm BCR}\sim -2k_{\rm bf}$, hence the wavevector of $\psi_{\rm BCR}$ in Eq.~\ref{eq:BdG1} becomes $k_{\rm bf}+\Delta k_{\rm BCR}\sim-k_{\rm bf}$.

The HR and p-mode spectra extend outward and inward from the BH, with positive and 
negative wavevectors, respectively. The wavevector spectra at $x_{\rm I}$(diagonal red line, left) and $x_{\rm O}$(diagonal blue, right) are shown on the right panel, with the modes labeled (except for $v^{*}_{\rm BCR}$) in the figure.

We compare the WFT spectra [Figs.~\ref{fig:STFT_wt}(b) and (c)] with the BdG dispersion relations [Fig.~\ref{fig:STFT_wt}(a)] at 
$x_{\rm I}$ and $x_{\rm O}$, corresponding to the intersections of the diagonal red (left) and diagonal blue (right) lines with the horizontal green line in Fig.~\ref{fig:M20}(b). The numerical values of $\Delta \omega$ and $\Delta k$ obtained from the WFT spectra of the GP solution are displayed in Table \ref{table:FT}, along with those predicted from the BdG dispersion relations.
The inputs to the BdG prediction are just (i) the assumption of zero frequency in the WH frame, and (ii) the velocity of the BH frame relative to the WH frame. The GP spectra and BdG predictions agree to within 5\%. Note that the flow is not perfectly stationary, so that the zero frequency of the initial BCR is not perfectly conserved. Also, the speed of the WH changes slightly over time, which gives rise to the uncertainty in $\Delta \omega_{\rm BdG}$ and $\Delta k_{\rm BdG}$.

\begin{table}[ht]
\caption{Numerical values of relative mode frequency $\Delta\omega$(rad/ms) and wavevector $\Delta k$($\mu$m$^{-1}$) from the GP Fourier spectra (FT) and from the WH-zero-frequency 
BdG dispersion relation (BdG). The uncertainty for the former is estimated by the widths of the Gaussians fitting the spectral peaks in Figs.~\ref{fig:STFT_wt}(b) and (c), and the uncertainty for the latter is due to the variation of the speed of WH.}
\label{table:FT}
\begin{tabular}{l c c c c}
\hline \hline
Modes   &  $\Delta\omega_{\rm FT}$ & $\Delta\omega_{\rm BdG}$   & $\Delta k_{\rm FT}$ &  $\Delta k_{\rm BdG}$ \\ \hline
$u_{\rm BCR}$ & -0.21(3) & -0.20$\pm0.01$ & 2.42(8) & 2.45$\pm0.02$ \\
$u_{\rm p}$   & -0.21(3) & -0.20$\pm0.01$ & 0.26(8) & 0.26$\pm0.01$  \\
$u_{\rm HR}$  &  0.20(3) & 0.20$\pm0.01$ & 0.65(12) & 0.65$\pm0.02$ \\
\hline \hline
\end{tabular}
\end{table}

\subsection{Hawking temperature}
\label{sec:temp}
The spontaneous emission 
from a black-hole horizon is thermal, with temperature $T_{\rm H}=\hbar\kappa/(2\pi k)$, where $\kappa$ is the surface gravity \cite{hawking1,hawking2}. 
In the sonic analog, the surface gravity becomes 
$\kappa = d(v+c)/dx$, evaluated at the horizon \cite{unruh1981}.
The Hawking mode of the stimulated radiation 
is excited with a coefficient $\beta$, and the partner mode 
with a coefficient $\alpha$, corresponding, in effect, to 
transmission and reflection coefficients. 
The ratio $|\beta/\alpha|$
carries the signature of the thermal 
prediction \cite{Unruh:1994je,Unruh:2014hua}, 
\begin{equation}\label{ratio}
\frac{|\beta|}{|\alpha|}=
\frac{|V_{\rm HR}/v_{\rm HR}|}{|U_{\rm p}/u_{\rm p}|}
=\exp(-\pi\Delta\omega/\kappa).
\end{equation}
Here $(U_{\rm p},V_{\rm HR})$ are the full mode amplitudes, which can be captured from the WFT spectra, 
and $(u_{\rm p},v_{\rm HR})$ are the normalized BdG amplitudes defined in Eq.~\ref{eq:uv}.

To test the thermal prediction we 
evaluate the mode amplitudes at a pair of points $x_{\rm p}$ and $x_{\rm HR}$
with a common retarded time, defined by phase velocity, at the BH.
These points are denoted by the magenta (left) and cyan (right) dots on the horizontal green line in 
Fig.~\ref{fig:M20}(b). The common retarded time on the horizon is $t=588$ ms, for which we
find the surface gravity $\kappa\sim350$ s$^{-1}$ (using $v$ and $c$ computed directly from the GP wavefunction, see Appendix~\ref{app:temperature}. The thermal prediction for
$\omega=200$ rad/s is $|\beta/\alpha|=0.17^{+0.05}_{-0.04}$, allowing for a 5\% 
uncertainty in $\omega$ and a 10\% uncertainty in $\kappa$. This agrees reasonably well with the ratio 0.21 computed directly from the amplitudes according to the thermal prediction. The Hawking temperature for the case M2 depicted in Fig.~\ref{fig:M20}(b) is $T_{\rm H}=0.43\,$ nK. The temperature equivalent of the chemical potential, $\mu$, in that case is $\mu/k=2.5\,$ nK.

Several factors could play a role in causing the GP ratio $|\beta/\alpha|$  to differ
from the thermal prediction. First, the latter is exponentially sensitive to the value of $\kappa$, so the time dependence of $\kappa$ can introduce
a significant effect. Second, phonon dispersion can produce deviations that depend on how large are $\kappa$ and $\omega$ compared to the sound speed over the healing length,
$c/\xi\sim 320{\rm s}^{-1}$, and on how wide is the linear regime of the function $c+v$ around the BH ($\sim$ 6 $\mu{\rm m}$), compared to $c/\kappa\sim$ 1.2 $\mu{\rm m}$ \cite{two_regimes}. And third, nonlinearity of the GP modes could lead to deviations from the linear prediction.

\section{Discussion}

\subsection{Comments on the lasing mechanism}

The results in this paper, together with those of \cite{correlations} which includes fluctuations, 
establish that the BH laser phenomenon is not present, or at least not significant, in our simulations of the experiment of Ref.~\cite{nphys3104}.
Instead, we have traced the standing wave and its growth to the BCR mechanism.  
But the question remains as to why lasing does not occur, given that 
the system exhibits a flow structure of the type that can lead to the laser instability.

We can suggest two factors that may lie behind this: one is that the growth rate of the instability may be too slow to 
show any significant growth during the timespan of the step sweep.
The other is that the time dependence of the flow structure and cavity size may lead to a detuning of the instability.

Regarding the time available for
the laser instability, what matters is both the number of ``cycles'' that occurs, and the amplification factor in each cycle. We estimate using the dispersion relation a ``round trip" time for modes propagating between the BH and WH of $\sim 50$ ms. This is consistent with Fig.~\ref{fig:stimulated_HR}(b), from which one can see that the p-mode that is created at the BH at $t\sim$ 60 ms [Fig.~\ref{fig:stimulated_HR}(b)] takes $\sim 30$ ms to reach the WH, and that the BCR takes $\sim 20$ ms to propagate from the WH to the BH. 
An initial partner mode excitation that reaches the left end of 
the cavity just after the WH has formed at $t\sim 20$ ms,
i.e., at the earliest possible time, would thus have time for no more than two return trips to the BH before $t=120$ ms.

Regarding the time-dependent flow, the black hole laser scenario was originally introduced, and has been studied, in the setting of stationary flow structure. In the experiment, however, the background density changes significantly as the step moves from the edge of the condensate to the center, changing the sound speed. In addition, the size of the supersonic cavity is time dependent due to the receding WH. This causes a repetitive Doppler shift ($\propto \Delta k \Delta v$) on the modes inside the cavity. The Doppler shift is greater for modes possessing a larger wavevector, for which the frequency shift in one cycle is comparable to the initial frequency. 
This is explained in more detail in Appendix~\ref{app:FT_regime_exp}, where the 
detuning effect is illustrated on the dispersion relation graph in Fig.~\ref{fig:FT_steinhauer}(c).
The upshot is that the effect of moving WH horizon cannot simply be treated as an adiabatic evolution of the static case.

An unstable mode responsible for the laser effect contains right-moving components with large wavevector, arising from mode conversion of the left-moving p mode at the WH horizon, which suffer a significant Doppler shift  relative to the frequency of the incoming p mode. When the cavity is small, the lasing can  be dominated by a single, fastest growing unstable mode \cite{Coutant:2009cu}. In that setting, the above time-dependent effects might ``detune" the laser, inhibiting the self-amplification mechanism. 

\subsection{Comparison with other simulated results}
\label{sec:comparison}

A study having some overlap with ours was reported by Tettamanti {\textit{et al.}} \cite{Tettamanti}. They established the hydrodynamic character of the experimental observations, and identified the Bogoliubov-\v{C}erenkov (BCR) mode as responsible for initiating the instability, both of which are consistent with our findings. Our accounts differ, however, regarding the subsequent evolution. They report that the resulting Hawking radiation is self-amplifying, and that the growing wave pattern between the horizons results from this amplification and the interference between counter-propagating waves. However, it is hard to glean from the paper on what basis that conclusion was drawn. 
The local wavevector spectrum in the standing wave region is shown in [Fig.~\ref{fig:exp}(b)],
and discussed in Sec.~\ref{sec:mechanism}. 
We find that the standing wave pattern is simply the result of interference between the BCR mode and the background flow, and grows
due to the growing condensate density. An additional, long wavelength, left-propagating partner mode of Hawking radiation is evident in the spacetime portrait [Fig.~\ref{fig:stimulated_HR}(b)]. However, this appears well after the standing wave has formed, and has much smaller amplitude. 

Another numerical investigation of this system, by Steinhauer and de Nova \cite{SR}, appeared while we were preparing revisions of our manuscript. They report findings indicating that the BCR component to the standing wave can not be generated at the WH horizon because it appears before the WH horizon forms. We suspect that this discrepancy with our findings may be traced to their use of a definition of the WH horizon that does not coincide with the location of the stationary, super-to-subsonic transition, as explained more fully in Sec.~\ref{sec:Locating}. 

In addition, Ref.~\cite{SR} argues that self-amplifying Hawking radiation can be distinguished from what they call the ``background ripple" by the presence of time dependence, since
the power in a monochromatic traveling wave such as the BCR would not oscillate in time. To exhibit the time dependence,
they evaluate a temporal Fourier transform of the spatial Fourier transform of the density, normalized by the square of the spatial average of the density in the cavity. This has a peak at a characteristic frequency which, they assert, is a signature of the self-amplifying Hawking radiation. 

We also find time dependence associated with the standing wave in the cavity, which can be seen in the oscillatory features in Fig.~\ref{fig:GP_growth}(a).
These oscillations result from interference with the Hawking partner mode, as can be seen in Fig.~\ref{fig:stimulated_HR}(b). This might be the source of the time dependence found in Ref.~\cite{SR}. The dominant period for this time dependence, which can be read from Fig.~\ref{fig:GP_growth}(a), is of order 40 ms. This corresponds to an angular frequency of order 0.16 rad/ms, i.e.,\  $\omega/\omega_{\rm max}\sim 0.2$ (where
$\omega_{\rm max}\sim 0.75$ rad/ms is the maximum allowed frequency in the cavity region), which is
not far from the values found in Ref.~\cite{SR}.

\section{Conclusion}
\label{sec:conclusion}
To conclude, we summarize the evidence that the black hole laser effect plays no role in our GP simulations. First, we find 
only one unstable mode, the BCR mode, which grows in proportion to the square of the background flow density.
Lasing action, by contrast, would have no reason to satisfy this relation. Second, this growing mode has zero frequency in the WH frame, which differs from the BH frame due to a nonzero relative motion of the two horizons. 
And, third, the windowed Fourier spectrum in the cavity reveals no other frequency components. The Hawking partner radiation has the same frequency as the BCR mode, as would be expected if it results from stimulation by the monochromatic BCR. 

Our simulations did not include quantum fluctuations capable of spontaneously producing Hawking radiation.
However, in a related paper \cite{correlations}
we have studied the effect of quantum fluctuations using the truncated Wigner approximation, and we found no evidence there of any mode growth beyond that found here. Moreover, the results of Ref.~\cite{correlations} reveal that the key features of the observed density correlation function can all be produced by modulation of the BCR standing wave caused by atom number variations and quantum fluctuations.

Finally, we investigated various regimes of potential experimental parameters, and found a regime where a
sharper signal of HR is obtained, and
in which a BdG mode description is valid. 
This enabled us to carry out a detailed quantitative check of our proposed mechanism, stimulation of the Hawking radiation by a Doppler shifted, zero frequency BCR standing wave. This enhanced parameter regime could provide a useful guide for future experimental investigations of stimulated Hawking radiation in this setting.

\section{Acknowledgements}
We thank R. Parentani for numerous helpful discussions and suggestions, and J. Steinhauer for stimulating correspondence and criticism.
This material is based upon work supported by the U.S.\ National Science Foundation Physics Frontier Center at JQI and
grants PHY--1407744, 
PHY--1004975 and PHY--0758111, and by the Army Research Office Atomtronics MURI. 

\appendix

\section{Characterization of the experimental condensates and description of simulation procedures}
\label{app:sim_details}

The condensate the in the experiment of Ref.~\cite{nphys3104}
is tightly confined in two transverse dimensions, and elongated in the third dimension, with a scale ratio $\sim 1:20$. It is thus approximately one-dimensional, and for all the simulations in this paper we have employed the one-dimensional description, except for some comparative simulations described in this appendix. Here we begin by describing a three-dimensional model of the system, and then proceed to explain the reduction to an effective, one-dimensional model. 
We also compare this 1D model to a 3D one, and find that the 1D model accurately captures the important features of the 3D dynamics.

The Gross--Pitaevskii (GP) equation, giving the mean field description of a BEC in three dimensions, takes the form
\begin{equation}
i\hbar\frac{\partial\Psi({\bf r},t)}{\partial t} = 
\left(-\frac{\hbar^{2}}{2m}\nabla^{2} + V({\bf r},t) + g_{\rm 3D}N\left|\Psi\right|^{2}\right)
\Psi({\bf r},t),
\label{tdgp3d}
\end{equation}
where $N$ is the number of condensate atoms, $m$ is the mass of a condensate atom, $g_{\rm 3D}=4\pi\hbar^{2}a/m$ where $a$ is the $s$--wave scattering length, and $V({\bf r},t)$ is the full external potential. In the case of the experiment, 
the potential is given by
\begin{equation}
V({\bf r},t) = U({\bf r}) + U_{\rm step}({\bf r},t).
\end{equation}
The potential in which the initial condensate was formed in the experiment of Ref.~\cite{nphys3104} is denoted by $U({\bf r})$, and $U_{\rm step}({\bf r},t)$ is the potential for the step that was swept along the length of the condensate, as shown schematically in Fig.~\ref{fig:exp} of our paper.  
The wave function for the initial condensate, $\Psi_{0}({\bf r})$, satisfies the time--independent GP equation: 
\begin{equation}
\left(-\frac{\hbar^{2}}{2m}\nabla^{2} + U({\bf r}) + g_{\rm 3D}N\left|\Psi_{0}({\bf r})\right|^{2}\right)
\Psi_{0}({\bf r}) = \mu_{0}\Psi_{0}({\bf r}),
\label{tigp3d}
\end{equation}
\noindent where $\mu_{0}$ is the chemical potential of the ground-state condensate. 

We begin by describing the most accurate 3D GP model,
given the information in Ref.~\cite{nphys3104}
about the experiment. We first found the potential, $U({\bf r})$, that is produced by the red--detuned trapping laser specified in Ref.~\cite{nphys3104}.  There, the laser beam characteristics are stated in terms of its wavelength, $\lambda = 812 $  nm, and the beam waist, $w_{0} = 5 $ mm.  We used these data to model the trapping laser light as a focused ideal Gaussian laser beam.  Thus the trapping potential is proportional to the beam intensity:
\begin{equation}
U({\bf r}) = U_{0}\left[1  -  
\left(\frac{w_0}{w\left(x\right)}\right)^2 
\exp\left(\frac{-2\rho^2}{w^2\left(x\right)}\right)\right].
\label{Intensity}
\end{equation}
where $x$ is the axis of light propagation, $\rho = \sqrt{y^2 + z^2}$ is the transverse (axial) radial coordinate, $w_0$ is the beam waist, $U_{0}$ is proportional to the peak laser intensity and
\begin{equation}
w(x) = w_0 \sqrt{1 + \left(\frac{x}{x_0}\right)^2} \, ,
\quad
{\rm where}
\quad
x_0 = \frac{\pi w_0^2}{\lambda}.
\label{wofz}
\end{equation}
We have chosen the origin of energy so that $U({\bf r})$ vanishes at the center of the trap, $U(0) = 0$.  
Reference~\cite{nphys3104} also states that, since the long axis of the needle--shaped condensate lies in a horizontal plane, the effect of gravity is mostly (all but 9\%) compensated for by an external magnetic field with a vertical gradient.  For simplicity, in our model we take the gravitational and compensating magnetic forces to cancel exactly. 

Specification of the wavelength and beam waist fixes all the parameters in $U({\bf r})$ except for $U_{0}$.  Reference~\cite{nphys3104} gives the axial trap frequency as $\omega_{\rho}/2\pi = 123$ Hz.  We used this frequency to determine $U_{0}$ by expanding $U({\bf r})$ to second order about ${\bf r}= 0 $ :
\begin{equation}\label{U(r)}
U({\bf r}) \approx
\left(\frac{2U_{0}}{w_{0}^{2}}\right)\rho^{2} + 
\left(\frac{U_{0}}{x_0^{2}}\right)x^{2} \equiv
\frac{1}{2}m\omega_{\rho}^{2}\rho^{2} +
\frac{1}{2}m\omega_{x}^{2}x^{2}.
\end{equation}
Thus, $U_{0} = (1/4)m\omega_{\rho}^{2}w_{0}^{2}\approx 39 \, k \, \mathrm{nK}$, where $k$ is the Boltzmann constant.  This completes our determination of $U({\bf r})$ from the experimental parameters. The full list of experimental parameters is given in Table~\ref{Parameters}.

\begin{figure}[htb]
\includegraphics[width=3.3in]{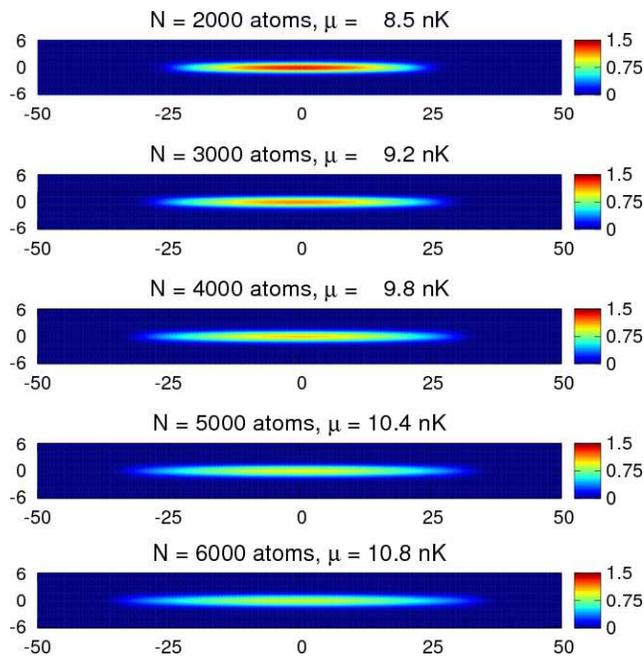}
\caption{Optical densities of the condensate ground states from 3D GP simulations for atom numbers $2000 < N < 6000$.  Each plot is also labeled with its associated chemical potential. Full horizontal and vertical scales are 100 and 12 micrometers, respectively, and the color box scale denoting optical density is graduated in arbitrary units.}
\label{initial_condensates}
\end{figure}

To determine the number of atoms in the condensate (which was not explicitly stated in Ref. \cite{nphys3104}) we simulate the initial condensate for different atom numbers $N$, as shown in Fig.~\ref{initial_condensates}. We find that  $N = 6000$ gives a best match of the axial length to that determined experimentally. Note that the chemical potential for $N = 6000$ is $\mu_{0}/k =10.8\,$ nK, which includes the radial kinetic and potential energies, and is greater than the reported experimental value, $\mu_{0}/k =8\,$ nK. It is not reported in \cite{nphys3104} how this value was determined;  however, for other quasi-low-dimensional BECs, a chemical potential usually refers to the maximal interaction energy determined by the maximal integrated density in the loosely-confined direction~\cite{1DBEC2001,pethick}.
In the simulated condensate with $N = 6000$, the maximal interaction energy in the axial direction is about 7.3 nK, which is comparable to the reported experimental value, 8 nK.

\begin{figure*}[htb]
\centering
\includegraphics[width=7in]{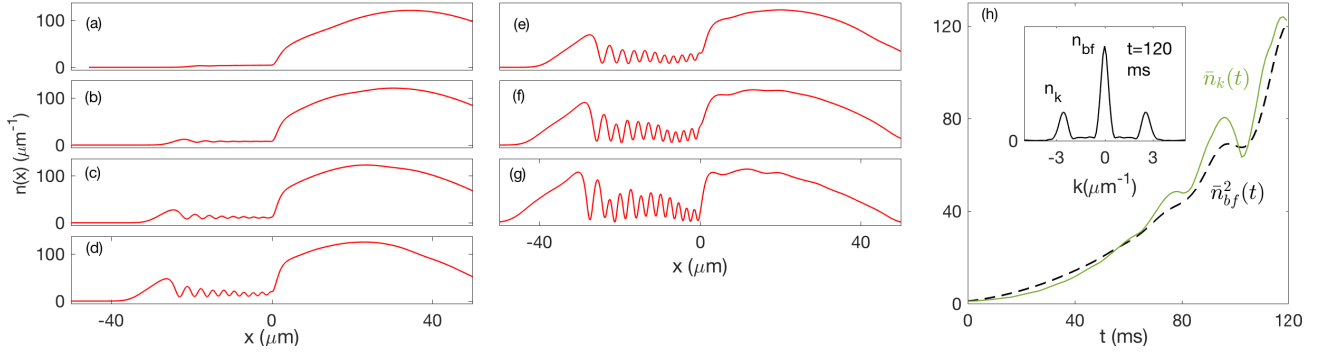}
\caption{\label{fig:GP2}
(a-g) Evolution of integrated density from a 3D simulation at 20 ms intervals with step $U_s/k=\text{ 5 nK}$, and viewed in the moving frame where $x=0$ defines the step edge. (h) Simulated growth of the standing-wave pattern in the supersonic region for $U_{s}/k =$  5 nK. Solid green: normalized standing-wave amplitude $\bar{n}_k(t)$, $\bar{n}_{k}(t)=n_{k}(t)/n_{k}(0)$, for which $\ln[\bar{n}_{k}(120)]\sim 4.8$. 
Dashed black: the square of background density, $\bar{n}_{\rm bf}(t)$, scaled to match the final standing-wave amplitude, $\bar{n}_{\rm bf}^2(t)=n_{\rm bf}^2(t)[\bar{n}_{k}(120)/n_{\rm bf}^2(120)]$. The growths of $n_{\rm bf}$ and $n_{k}$ are determined from a spatial WFT of $n(x)$ at $x=-12.5$ $\mu$m.}
\end{figure*}

\begin{figure*}[htb]
\centering
\includegraphics[width=7in]{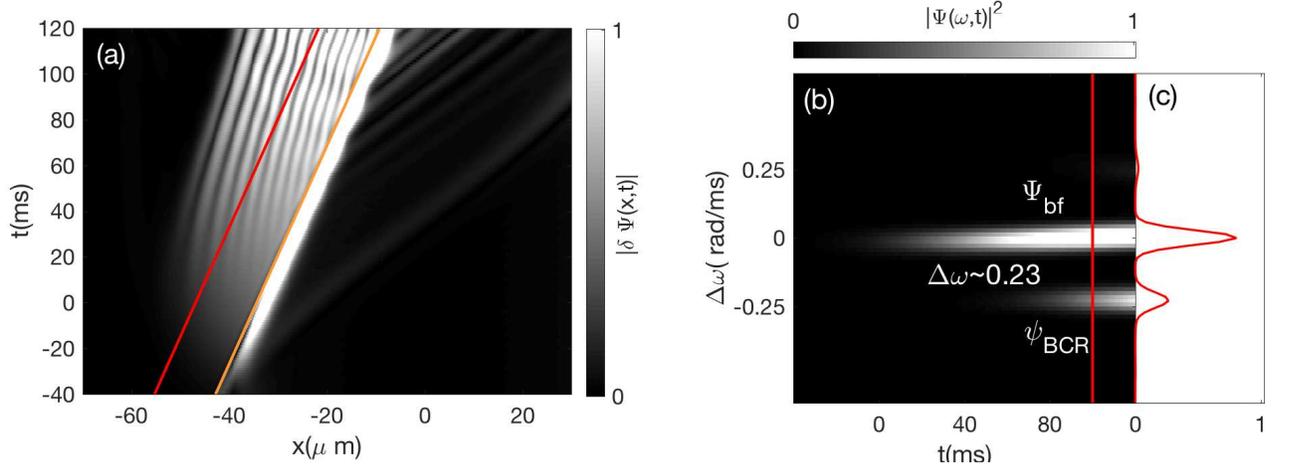}
\caption{\label{fig:FT_3d} Spacetime diagram and WFT frequency spectrum for the 3D simulation. Panel (a): time evolution of $|\delta\Psi(x,t)|$. Panel (b): windowed Fourier spectrum evaluated along the diagonal red (left) line in (b); panel (c) is the cut-through of the spectrum at $t=100$ ms. Note that the Doppler-shifted frequency $\Delta\omega_{\rm pair}\sim 0.23$ rad/ms, which is about twice the value from the 1D GPE.}
\end{figure*}

\begin{table}[ht]
\caption{Parameters of the trapped BEC as reported in, or inferred from (*), Ref. \cite{nphys3104}.  Uncertainties are not stated in Ref. \cite{nphys3104}, and we do not attempt to estimate them in this work.}  
\label{Parameters}
\begin{tabular}{l c r}
\hline \hline
Parameter & Value & Units\\ \hline
atom & $^{87}$Rb & \\
atomic state & $F=2, M_F =2$ & \\
trapping laser wavelength $\lambda$	&	812	& nm\\
beam waist $w_0$	&	5 & micron	\\
radial trap frequency $\nu$ & 123 & Hz\\
transverse energy level spacing $E$ & 6 & nK $k$\\
healing length $\xi$ & 2 & micron \\
nominal chemical potential  $\mu$ &  8 & nK $k$\\
*actual chemical potential $\mu$ &  10.4 & nK $k$\\
*axial length scale $x_0$ from eq. (\ref{wofz}) & 97 & micron \\
*number of condensate atoms $N$ & 6000 & atoms\\
\hline \hline
\end{tabular}
\end{table}

\subsection{Reduction to a one-dimensional system}
\label{app:1dGPE}
When a condensate is tightly confined in the radial direction, and the integrated density $n$ in the axial direction satisfies $na\ll1$, it can be viewed as quasi-one-dimensional \cite{1DGPE,stringari}.
For the experimental configuration, 
the axial to radial frequency ratio is given by 
$\omega_x/\omega_\rho= w_0 / (\sqrt{2}x_0) =4.5 \, \mathrm{Hz}/123 \, \mathrm{Hz}$,
which is much smaller than unity.
This shows that the system is tightly-confined in the radial direction.
Using the maximum of the integrated density from the 3D simulation, $n_{\rm max}\sim120$ $\mu{\rm m}^{-1}$, we estimate that the integrated 1D density satisfies 
$an_{\rm max}\approx 0.65$, which is less than, but not much smaller than, unity.  It is therefore not guaranteed that the 1D system is truly quasi-one dimensional. However, as discussed in Sec.~\ref{app:3dGPE}, we compared the 3D and 1D simulations and found the essential features to be quite similar.

To implement the simplest sort of 1D reduction, we approximate the wavefunction in the radial direction by the solution of a harmonic oscillator, so that
\begin{equation}
\Psi(\mathbf{r},t) = \frac{\exp\left[-\rho^2 /\left(2 d^2\right)\right]}{d\sqrt{\pi}}\Psi_{\rm 1D}(x,t),
\end{equation} 
where $d = \sqrt{\hbar/\left(m \omega_{\rho}\right)}$. Integrating the 3D GP equation over the Cartesian coordinates $y$ and $z$, we obtain a 1D GP equation with an effective interaction coefficient $g_{\rm 1D}=g_{\rm 3D}m \omega_{\rho}/h$:
\begin{eqnarray}
i\hbar\frac{\partial\Psi_{\rm 1D}(x,t)}{\partial t} &=& 
\left(-\frac{\hbar^{2}}{2m}\frac{\partial^2 }{\partial x^2} + V(x,t)\right)\Psi_{\rm 1D}(x,t)\nonumber\\
&+& g_{\rm 1D}N\left|\Psi_{\rm 1D}(x,t)\right|^{2} \Psi_{\rm 1D}(x,t),
\label{tdgp1d}
\end{eqnarray}
where $h$ is the Planck constant and $V(x,t)$ is the full external potential $V(\bf{r},t)$ evaluated at $y=z=0$. We define the coefficient of the nonlinear term as, $g=g_{\rm 1D}N$, which is used extensively in the paper. 

We take the number of atoms determined by the 3D GP equation, $N=6000$, and use the 1DGPE [Eq.~\ref{tdgp1d}] to simulate the step-sweeping experiment. Note that Ref.~\cite{Tettamanti} simulated the  dynamics by using a 1D nonpolynomial nonlinear Schrodinger equation (NPSE) \cite{1DNPSE}, which incorporates the effect of a variable axial density on the transverse shape of the GP wave function (under the condition that the axial derivative of the transverse wave function is much smaller than the transverse derivative). Here we use a simpler 1D GP equation, which assumes a fixed transverse shape of the wave function. In the paper, and in other sections of the Appendix, we drop the subscript ``1D'' in $\Psi_{\rm 1D}(x,t)$ when referring to the 1D GP wavefunction.

\subsection{Solution of the time--dependent 1D GP equation}
\label{app:1dTDGPE}

The time-dependent 1D GP equation is solved by using the split-step Crank-Nicholson algorithm \cite{adhikari} on a 1D spatial grid of 320 $\mu$m with 4800 points, first propagating in imaginary time to obtain the initial stationary condensate, then propagating in real time with the given initial state to simulate the dynamics. To simulate the step-sweeping experiment \cite{nphys3104}, we use a step potential $U_{\rm step}(x,t)$, which takes the form
\begin{equation}
\label{eq:step}
U_{\rm step}(x,t)=-U_{\rm s}\Theta(x_{\rm s}(t)-x),
\end{equation}
where $\Theta$ is the Heaviside step function, $U_{\rm s}$ is the step strength, which takes the values of $U_{\rm s}/k=$ 3 nK and 6 nK, and $x_{\rm s}(t)$ represents the step location, moving at a constant speed, $v_s=$ 0.21 mm/s.

\subsection{Growing standing wave, spacetime portrait, and frequency spectrum from a 3D simulation}
\label{app:3dGPE}
To test the accuracy of the 1D simulation, we also simulated the  step-sweeping experiment using the 3D GP equation, with the potential \eqref{Intensity}, assuming the 
condensate shares the axial symmetry of the potential.
Figures~\ref{fig:GP2}(a)-(g) show the integrated density profiles with a potential step, $U_{\rm s}/k=$ 5 nK, which is adjusted slightly to match the cavity size with the experiment. The growth of the standing-wave amplitude, $\bar{n}_k$, and that of the background density, $\bar{n}_{\rm bf}$, are shown in Fig.~\ref{fig:GP2}(h). The standing wave grows by $\sim\exp(4.8)$, which is greater than in the 1D simulation 
($\sim\exp(4.4)$), but the growth relation, $\bar{n}_{k}\propto\bar{n}^2_{\rm bf}$, is preserved in the 3D simulation. 

Similarly, we calculate the spacetime portrait and the local frequency spectrum using the GP wavefunction at the center of the radial trap, $\rho=0$. The spacetime portrait in Fig.~\ref{fig:FT_3d}(a) shows very similar features as those in the 1D simulation, including the standing wave parallel to the WH, and the stimulated Hawking pair. The WH recession can also be seen in the portrait, which gives rise to a Doppler-shifted BCR frequency in the WFT spectrum in Fig.~\ref{fig:FT_3d}(b) and (c), $\Delta\omega\sim 0.23$ rad/ms. 

Although there are some quantitative differences with the 1D simulation, all the qualitative features found in the 1D GPE are preserved here: (i) the growth relation between the standing wave and the background density, (ii) the stimulated HR pair by the BCR, and (iii) the Doppler shift due to the WH recession.

\section{Windowed Fourier transform}
\label{app:FT}
Here we summarize the basic ideas of the windowed Fourier transform (WFT), and explain our use of it. In \ref{app:FT_definition}, we give the definition of WFT used here, and provide a few basic examples to show how it can resolve spectral information on non-stationary phenomena. In \ref{app:FT_vc}, we describe the application of the WFT to the determination of flow and sound speeds, $v(x)$ and $c(x)$, in inhomogeneous media. In \ref{app:FT_spectral_analysis}, we discuss calculations of the wavevector and frequency spectra displayed in Figs.~\ref{fig:STFT_wt}(b) and (c), and the additional spectra that distinguish the partner and BCR modes. In \ref{app:FT_regime_exp}, we show the windowed frequency spectrum for the experimental regime, and a comparison with the dispersion relation.

\subsection{Definition and examples}
\label{app:FT_definition}

A windowed Fourier transform \cite{gomes99fourier} $f(k,x)$ of a function $f(x)$ is defined as:
\begin{eqnarray}
\label{eq:FTkx1}
f(k,x)= \int^{\infty}_{-\infty} dy\, f(y)w(y-x;D)e^{-iky},
\end{eqnarray}
where $w(y-x;D)=\exp (-(y-x)^2/D^2)/\left(\sqrt{\pi}D\right)$ is a Gaussian window function of width $D$. With the filtering of the window, the transformed function $f(k,x)$ constitutes a local Fourier transform of $f(x)$, capturing features that vary on length scales much smaller than $D$.  For a plane wave with wavevector $q$ and amplitude $f_{q}$ , $f(x)=f_{q} \exp (iqx)$, the transformed function is $f(k,x)=f_{q} \exp (-(k-q)^2(D/2)^2)$ : a Gaussian in $k$-space, centered at $k=q$ with width $2/D$ and peak amplitude $f_{q}$. 

Suppose now that $f(x) = f_{q}(x) \exp(iqx)$, where $f_q(x)$ has weak dependence on $x$, and can be adequately approximated near a point $x_0$ by
\begin{eqnarray}
\label{eq:FTkx2}
f_{q}(x)= f_q(x_0)+f_q^{\prime}(x_0)(x-x_0).
\end{eqnarray}
Then for sufficiently small values of $D$, the WFT of $f(x)$ near $x=x_0$ is approximately
\begin{eqnarray}
\label{eq:FTkx3}
f(k,x_0)&\approx & f_q(x_0)e^{-(k-q)^2(D/2)^2} \nonumber\\
&+&f_q^{\prime}(x_0) \,i\frac{k-q}{2}D^2\, e^{-(k-q)^2(D/2)^2}.
\end{eqnarray}
Note that the second term vanishes at the peak position $k=q$, so that $f(q,x_0)\approx f_q(x_0)$. 

Finally, let $f(x)$ be composed of a number of such slowly--varying modes, 
\begin{eqnarray}
\label{eq:FTkx4}
f(x)= \sum\limits_{n} f_{q_n}(x)e^{iq_{n}x}, 
\end{eqnarray}
so that Eq.~\ref{eq:FTkx3} becomes
\begin{eqnarray}
\label{eq:FTkx5}
f(k,x_0)& \approx &  \sum\limits_{n} \left[f_{q_n}(x_0)+f_{q_n}^{\prime}(x_0)\left(i\frac{k-q_n}{2}D^2\right)\right]\nonumber\\ 
&&~~~\times e^{-(k-q_n)^2(D/2)^2}.
\end{eqnarray}
In $k$--space, each mode presents a Gaussian distribution centered on its respective $q_n$, whose peak value of $f(q_n,x_0)$ defines the local mode amplitude.  This is how we make quantitative determinations of the mode amplitudes that are discussed in our paper.

\subsection{Determination of the profiles of flow speed and the speed of sound}
\label{app:FT_vc}

\begin{figure*}[htb]
\includegraphics[width=7in]{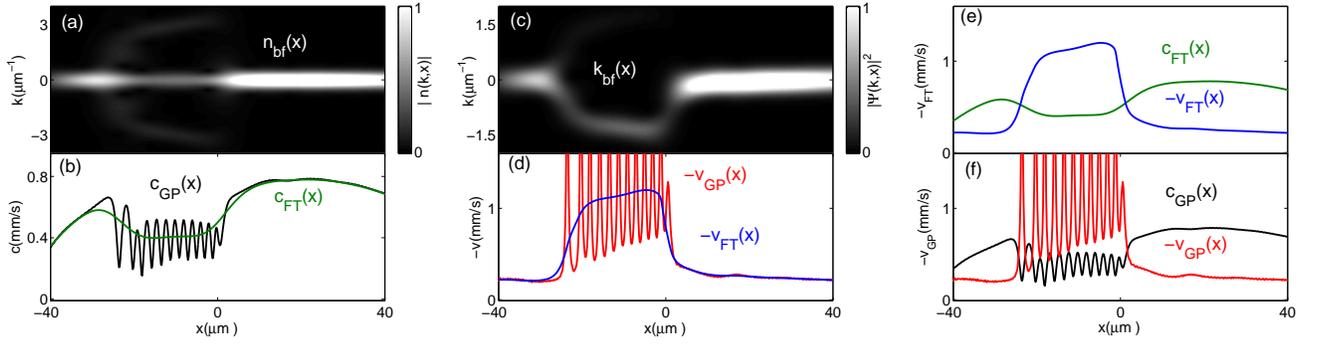}
\caption{Determination of flow speed $v(x)$ and the speed of sound $c(x)$. (a) Windowed wavevector spectrum of density $n(x)$ at $t_0=$ 80 ms [Fig.~\ref{fig:GP1}(e) in the paper]; (b) black: local speed of sound $c_{\rm GP}$ from total density $n(x)$; green (light gray): speed of sound $c_{\rm FT}$ from the background flow density $n_{\rm bf}(x)$, filtered with the Fourier spectrum; (c) windowed wavevector spectrum of GP wavefunction $\Psi(x)$; (d) red (gray, oscillatory): windowed flow speed, $v_{\rm GP}$, from a direct calculation on $\Psi(x)$; blue (dark gray, non-oscillatory): flow speed $v_{\rm FT}$ from the background flow wavevector $k_{\rm bf}(x)$; (e), (f) flow structure determined by the Fourier spectra (e) and that directly obtained from the GP wavefunction (f).}
\label{fig:vc_supplement}
\end{figure*}

As shown in the paper, during the sweep of the step, the time-dependent GP wavefunction $\Psi(x,t)$ exhibits excitation modes on top of the background condensate. To calculate the speed of sound $c(x)$ and flow speed $v(x)$ associated with 
the background condensate,
we extract the condensate from the full GP wavefunction with 
the help of a WFT.

First, the amplitude of the background flow at a given time $t_0$ can be calculated by applying a spatial WFT on the GP density $|\Psi(x,t_0)|^2 =n(x)$ , where $t_0$ is suppressed for brevity.
\begin{eqnarray}
n(k,x)= \int^{\infty}_{-\infty}dy \, n(y) w(y-x;D) e^{-iky}.
\end{eqnarray}
Figure~\ref{fig:vc_supplement}(a) shows the result of a spatial WFT of density $n(x)$ (which corresponds to Fig.~\ref{fig:GP1}(e) in the paper) with width $D=5$ $\mu$m. The central streak at $k\sim 0$ corresponds to the background flow, whose peak value gives rise to the background density $n_{\rm bf}(x)=|n(k\sim0,x)|$, as shown in Fig.~\ref{fig:vc_supplement}(b). The local speed of sound can then be expressed as by $c(x)=\sqrt{gn_{\rm bf}(x)/m}$. In Fig.~\ref{fig:vc_supplement}(b), we see that WFT method is appropriate in the slowly varying regions away from the two horizons: in the exterior region, $n_{\rm bf}(x)$ matches the GP density $n(x)$; in the interior region, $n_{\rm bf}(x)$ is at about the average value of the density
oscillations.  Near the event horizons, on the other hand, the background density changes rather quickly, so that WFT introduces an unwanted averaging. In these regions, it is more appropriate to use the local GP wavefunction directly to define $v(x)$ and $c(x)$, since there are no significant excitations on the background flow, and the definition is strictly local.

Second, the flow velocity can be calculated by a WFT of the GP wavefunction 
\begin{eqnarray}
\label{eq:FTkx_n}
\Psi(k,x)= \int^{\infty}_{-\infty} dy\, \Psi(y)w(y-x;D)e^{-iky},
\end{eqnarray}
where again $t_0$ is suppressed for brevity. Figure~\ref{fig:vc_supplement}(c) shows the windowed wavevector spectrum $|\Psi(k,x)|^2$ with width $D=5$ $\mu$m. The dominant streak is the background flow, whose peak location $k_{\rm bf}(x)$ gives rise to the flow velocity in the rest frame of the step, $-v(x)=\hbar k_{\rm bf}(x)/m-v_s$, as shown in Fig.~\ref{fig:vc_supplement}(d); the peak value of the streak also corresponds to the background density $n_{\rm bf}(x)=|\Psi(k_{\rm bf},x)|^2$. In addition, we calculate the velocity profile by using the full GP wavefunction, $v_{\rm GP}(x)=\hbar/(m n(x)) \text{ Im} \left[ \Psi^{*}(x)d\Psi(x)/dx\right]-v_s$. We can see that WFT works well in regions apart from the event horizons, and effectively projects out the spatial oscillation present in $v_{\rm GP}(x)$.

Figures~\ref{fig:vc_supplement}(e) and (f) compare $v(x)$ and $c(x)$ from the windowed spectra [Fig.~\ref{fig:vc_supplement}(e)] with those obtained from the full GP wavefuntion [Fig.~\ref{fig:vc_supplement}(f)]. In short, near the event horizon, the approach of directly adopting the GP wavefunction gives more accurate speed profiles, with the correct horizon locations and the respective surface gravity; yet away from the horizons, the WFT effectively removes excitations from the background flow, and hence gives a more suitable definition for $v(x)$ and $c(x)$. 

\subsection{Spectral analysis with windowed Fourier transform}
\label{app:FT_spectral_analysis}

The spectral properties of excitation modes can be obtained by performing spatial and temporal WFTs on the condensate wavefunction. Given a GP wavefunction, $\Psi(x,t)$, we calculate its local wavevector spectrum and frequency spectrum by applying the WFTs. To obtain a local wavevector spectrum, we perform a spatial WFT on the wavefunction $\Psi(x,t_0)$ using Eq.~\ref{eq:FTkx_n} at a time $t_0$ in which excitation modes are present. The result is presented in Fig.~\ref{fig:STFT_wt}(c) in the paper, in which the excitation modes are resolved in addition to the background flow. Note that for the region on the right-hand side of the step, we perform the WFTs on the variation function $\delta\Psi(x,t)$ rather than $\Psi(x,t)$  (see Appendix~\ref{app:separation}), in order to subtract the background component and bring out the excitation mode in that region.

\begin{figure*}[htb]
\centering
\includegraphics[width=7in]{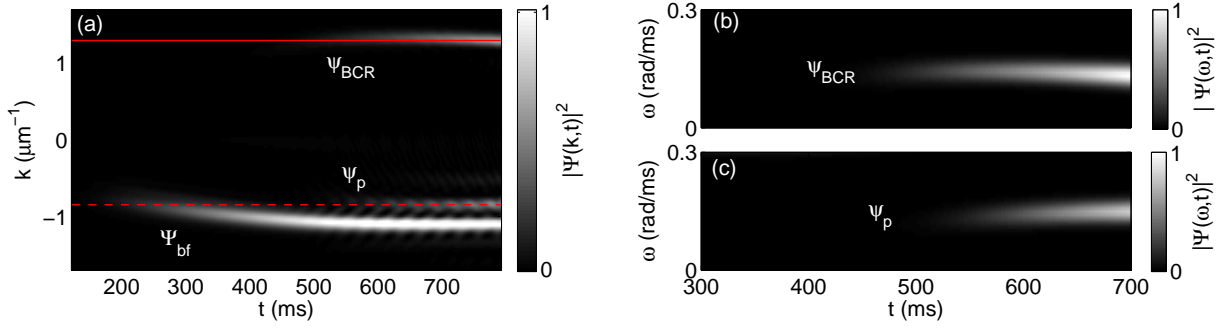}
\caption{\label{fig:bcr_p_spectra}Distinguishing the BCR mode ($\psi_{\rm BCR}$) and the partner mode ($\psi_{\rm p}$). (a) Local wavevector spectrum evaluated at $x_{\rm I}(t)$ for various times, in which the BdG modes ($\psi_{\rm BCR}$ and $\psi_{\rm p}$) of the same frequency are separated at different $k$ values; (b) the frequency spectrum of $\Psi(k_{\rm BCR},t)$, where $k_{\rm BCR}$ is indicated by the solid red line in (a); (c) the frequency spectrum of $\Psi(k_{\rm p},t)$, with $k_{\rm p}$ indicated by the dashed red line in (a).}
\end{figure*}

For a local frequency spectrum, we apply a temporal WFT at position $x_0(t)$ moving at constant speed $v_{\rm s}$ with the potential step:
\begin{eqnarray}
\label{eq:FTwt_n}
\Psi(\omega,t)= \int^{\infty}_{-\infty} d\tau\, \Psi(x_0(\tau),\tau)w(\tau-t;T)e^{i\omega \tau},
\end{eqnarray}
where $w(\tau-t;T)$ represents a Gaussian window function of width $T$, $w(\tau-t;T)=e^{-(\tau-t)^2/T^2}/\sqrt{\pi}T$; $x_0$ is selected to be both inside ($x_{\rm I}$) and outside the BH cavity ($x_{\rm O}$), which is indicated by the red (left diagonal) and blue (right diagonal) lines in Fig.~\ref{fig:M20}(b). The result is presented in Fig.~\ref{fig:STFT_wt}(b). In the figure, there are two modes ($\psi_{\rm p}$ and $\psi_{\rm BCR}$) overlapped in the frequency spectrum ($\omega\sim0.15$ $\mu$m) evaluated at position $x_{\rm I}(t)$. To resolve the two modes, we perform a spatial WFT evaluated at $x_{\rm I}(t)$  for various times
\begin{eqnarray}
\label{eq:FTkt_n}
\Psi(k,t)= \int^{\infty}_{-\infty} dy\, \Psi(y,t)w(y-x_{\rm I}(t);D)e^{-iky}.
\end{eqnarray}
The result is presented in Fig.~\ref{fig:bcr_p_spectra}(a), from which $\psi_{\rm BCR}$ and $\psi_{\rm p}$ are separated at different $k$ values, $k_{\rm BCR}$ (solid red line) and $k_{\rm p}$ (dashed red line). Furthermore, by performing a temporal WFT on $\Psi(k,t)$ at the two wavevectors, we resolve the overlapped streaks in the initial frequency spectrum at $\omega\sim0.15$ rad/ms, as shown in Figs.~\ref{fig:bcr_p_spectra}(b) and (c).

\subsection{Windowed frequency spectrum and dispersion relation for the experimental regime}
\label{app:FT_regime_exp}
Here we present the frequency spectrum for the experimental regime, in comparison with the prediction from the BdG dispersion relation as in Sec.~\ref{sec:BdG}. This shows that no black hole laser effect is apparent in our simulation of the experiment of Ref. \cite{nphys3104}.

\begin{figure*}[htb]
\centering
\includegraphics[width=7in]{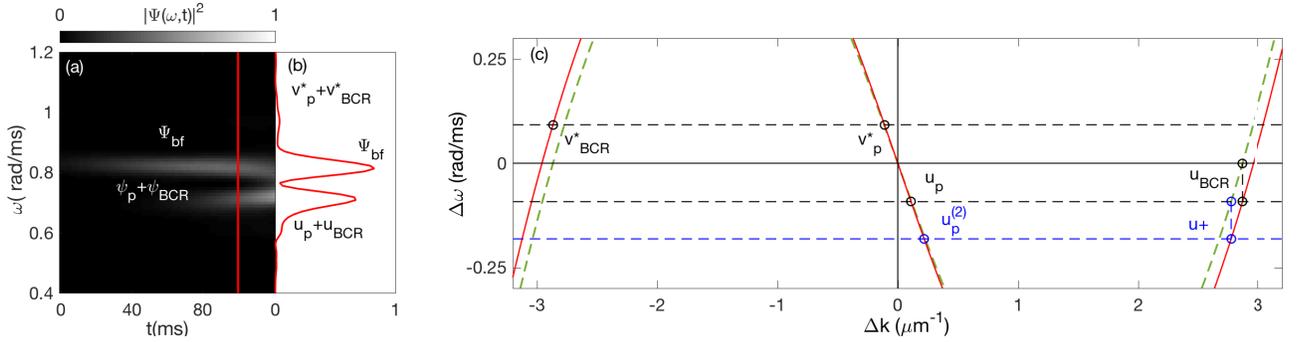}
\caption{\label{fig:FT_steinhauer} Windowed frequency spectrum and dispersion relation for the experimental regime. (a) Frequency spectrum evaluated at $x=x_{BH}-12$ $\mu$m. (b) Cut-through along the vertical red line in (a). (c) Dispersion relations in the WH (dashed green curve) and BH (solid red curve) reference frames, evaluated at $t=$ 100 ms.  The dashed black lines show the frequencies of the BCR mode in the BH frame, which stimulates the first HR pair.
The horizontal dashed blue (bottom) line indicates the frequency of a positive-norm mode $\psi_+$ in the BH frame, which stimulates the 
second HR pair. Note that $u_{+}$ represents the $u$-component of $\psi_{+}$ (Eq.~\ref{eq:BdG1}); $u_{\rm p}$ and $u_{\rm p}^{(2)}$ denote the $u$-components of the first and the second p-modes, respectively. }
\end{figure*}

We apply the temporal WFT on $\Psi(x,t)$ at a position about the center of the cavity, $x_{\rm I}=x_{BH}-12$ $\mu$m, indicated by the diagonal red (left) line in Fig.~\ref{fig:stimulated_HR}(b). The resulting frequency spectrum is given in Figs.~\ref{fig:FT_steinhauer}(a) and (b).
The streak that appears from early times shows the frequency of the background flow wavefunction $\Psi_{\rm bf}$.
The lower streak corresponds to the superposition of the BCR and the partner mode. The cut-through at $t=$ 100 ms is shown in Fig.~\ref{fig:FT_steinhauer}(b), from which can be seen the relative frequency (of the $u$-components), $\Delta\omega\sim $ -0.11(3) rad/ms.

We also predict this relative frequency using the dispersion relation, as in Fig.~\ref{fig:STFT_wt}(a) for the enhanced regime. The assumption that the BCR is the zero-frequency mode in the WH frame determines $\Delta k_{\rm BCR}=2.9$ $\mu$m$^{-1}$. Taking into account the velocity difference between the WH and the BH, $\Delta v\sim 0.03$ mm/s, the relative frequency of BCR (and p-mode) in the BH frame is given by $\Delta\omega=-\Delta k_{\rm BCR} \Delta v\sim - 0.09$ rad/ms (for the components $u_{\rm BCR}$ and $u_{\rm p}$), which is indicated by the lower dashed black line in Fig.~\ref{fig:FT_steinhauer}(c). This predicted frequency is within the uncertainty of the measured WFT value.

The wavelength of the partner mode predicted using the dispersion relation is $\lambda_{\rm p}\sim 57$ $\mu$m, which is greater than the width of the supersonic cavity $L\sim 25$ $\mu$m. Therefore the partner cannot be treated in the WKB approximation,
and the discrete spectrum of cavity modes modifies the emission, unlike in the M2 regime where the ratio $\lambda_{\rm p}/L$ is smaller. This may explain the irregular wavelength of the HR in the experimental regime seen in Fig.~\ref{fig:stimulated_HR}(b).

When the p-mode scatters at the WH, it creates a pair of positive-norm ($\psi_{+}$) and negative-norm ($\psi_{-}$) modes \cite{nphys3104} [here we only show the former, $u_{+}$ in Fig.~\ref{fig:stimulated_HR}(c)], whose frequency is the same as that of the partner in the WH frame.
Due to the relative velocity between BH and WH, $u_{+}$ has a shifted frequency in the BH frame [the horizontal blue (bottom) line in Fig. \ref{fig:FT_steinhauer}(c)], lower than the frequency of the first p-mode,  
and it stimulates the second p-mode ($u_{p}^{(2)}$) at that shifted frequency. The repetitive scatterings at the horizons therefore do not occur at a single frequency, as they would in the static case, i.e.,\
with zero WH horizon velocity. For the BCR, it can be seen in Fig.~\ref{fig:FT_steinhauer}(c)
that the frequency shift per cycle is comparable to the frequency itself, $\Delta \omega/\omega \sim 1$. The motion of the WH is therefore not well within the adiabatic regime. Hence, for large wavevectors, like those of the BCR, the static analysis of the black-hole lasing phenomenon is not reliable for predicting what happens with the moving WH horizon.

\section{Separation of fast and slow oscillation of condensate wavefunction}
\label{app:separation}

To separate the HR from the subsonic background flow, we apply a smoothing procedure to separate fast oscillatory modes from the slowly-varying components in the GP wavefunction. The procedure is equivalent to calculating the moving average of a discrete data set, which smooths out short-term fluctuations. Here, the moving average of wavefunction $\Psi(x)$ is defined as
\begin{eqnarray}
\label{eq:smoothing}
\bar{\Psi}(x)=\frac{1}{2D} \int^{x+D}_{x-D} dy\, \Psi(y),
\end{eqnarray}
where the integral serves as a square window of width $2D$ centered at $x$, over which $\Psi(x)$ is being averaged. For components in $\Psi(x)$ with wavelength much shorter than $D$ (i.e., $D \gg\pi/k$), the integral would give rise to an average of zero, leaving those that are slowly varying in space (i.e., $\pi/k \gg D$) in $\bar{\Psi}(x)$. 

\begin{figure*}[htb]
\includegraphics[width=7in]{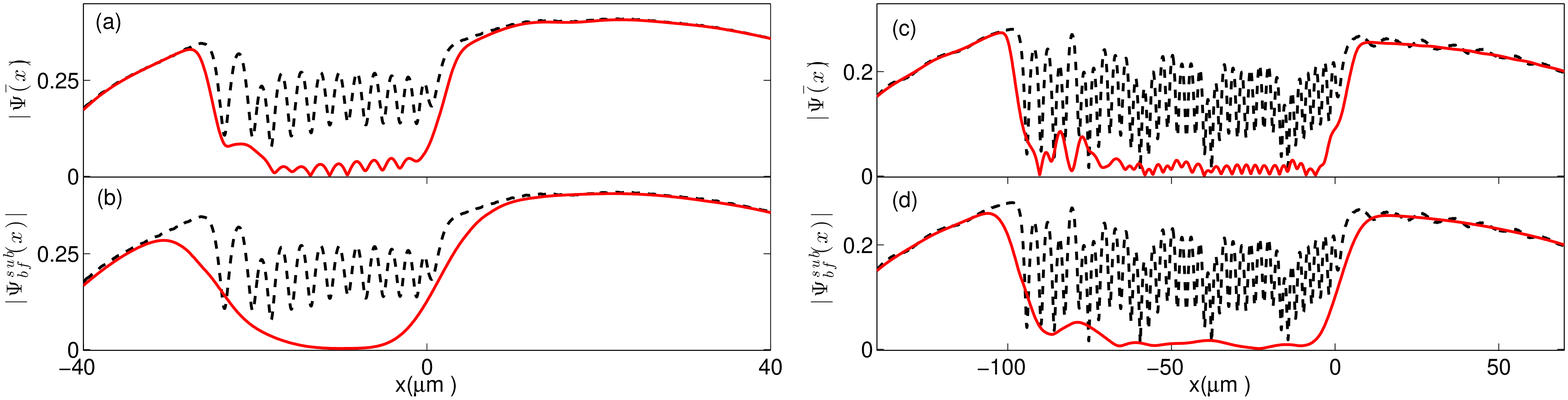}
\caption{Comparison of the smoothed wavefunction [(a), (c)], $\bar{\Psi}$, with subsonic background flow, $\Psi_{\rm bf}^{\rm sub}$, obtained by a spatial WFT [(b), (d)]. Experimental regime [Fig.~\ref{fig:GP1}(e) in the paper]: (a) smoothed wavefunction using Eq.~\ref{eq:smoothing} with window width $D=5.4$ $\mu$m; (b) spatial WFT $|\Psi(k,x)|$ with Gaussian width $D=5$ $\mu$m evaluated at $k\sim 0$. Modified regime [Fig.~\ref{fig:M20}(a) in the paper]: (c) smoothed wavefunction using Eq.~\ref{eq:smoothing} with window width $D=11.4$ $\mu$m; (d) spatial WFT $|\Psi(k,x)|$ with Gaussian width $D=7$ $\mu$m evaluated at $k\sim 0$. Note that the dashed black curve in all the panels corresponds to the GP wavefunction, $|\Psi(x)|$.}
\label{fig:psi0_supplement}
\end{figure*}

\begin{figure*}[htb]
\includegraphics[width=7in]{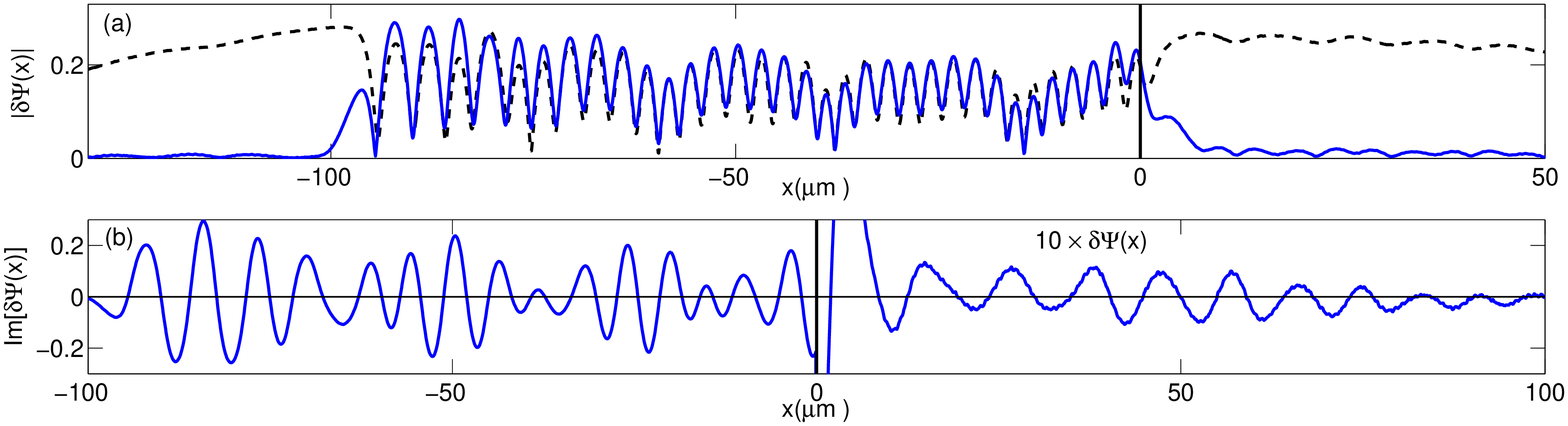}
\caption{Variation function in the M2 regime, $\delta\Psi(x)$, calculated with window width $D=$ 11.4$\mu$m. (a) Solid blue: $|\delta\Psi(x)|$; dashed black: $|\Psi(x)|$. (b) The imaginary part of $\delta\Psi(x)$, whose value on the right-hand side of the step (denoted by the vertical black line) is multiplied by 10. }
\label{fig:delta_psi_supplement}
\end{figure*}

According to Fig.~\ref{fig:vc_supplement}(c), the background flow in the subsonic region ($\Psi_{\rm bf}^{\rm sub}$) has $k\sim0$ and can be separated from the GP wavefunction through Eq.~\ref{eq:smoothing}, such that $\bar{\Psi}(x) \approx \Psi_{\rm bf}^{\rm sub}(x)$. Furthermore, highly oscillatory components in the wavefunction, including all the excitation modes ($\psi_j$) and the supersonic background flow ($\Psi_{\rm bf}^{\rm sup}$), can be obtained by subtracting the GP wavefunction with the non-oscillatory component, $\delta\Psi(x) = \Psi(x)-\bar{\Psi}(x)$. Thus, the variation $\delta\Psi(x)$ can be expressed as
\begin{eqnarray}
\label{eq:delta_psi}
\delta\Psi \approx \psi_{\rm p}+\psi_{\rm HR}+\psi_{\rm BCR}+\Psi_{\rm bf}^{\rm sup}.
\end{eqnarray}
Figs.~\ref{fig:psi0_supplement}(a) and (c) show the application of Eq.~\ref{eq:smoothing} to obtain a smoothed profile of $|\bar{\Psi}(x)|$ in the experiment of Ref. \cite{nphys3104}, and for a simulation in the M2 enhanced regime.  A separate calculation using the spatial WFT is shown in Figs.~\ref{fig:psi0_supplement}(b) and (d), in which $|\Psi_{\rm bf}^{\rm sub}|$ is evaluated by taking the peak amplitude $|\Psi(k,x)|$ at $k\sim0$. Both approaches agree with the GP wavefunction at regions away from the event horizons, capturing the background component outside the supersonic cavity. This gives rise the variation function $\delta\Psi(x)$, which nicely approximates the components in Eq.~\ref{eq:delta_psi}. Figure~\ref{fig:delta_psi_supplement} shows the variation $\delta\Psi(x)$ for the M2 regime, which agrees with the GP wavefunction inside the BH cavity. Note that in Fig.~\ref{fig:delta_psi_supplement}(b) we have multiplied $\delta\Psi(x)$ by a factor of 10 for $x>x_{\rm BH}$ to bring out the HR mode in the exterior region. 

\section{BEC parameter regimes in which Hawking radiation has greater visibility}
\label{app:parameter}

To find a more distinctive signature of HR, we study the GP evolution in different parameter regimes where the frequency of the trapping potential, $\omega_{x}$, and the depth, $U_{\rm s}$, and speed, $v_{\rm s}$, of the potential step are varied away from the values ($\omega^{0}_x$, $U_{\rm s}^{0}$, $v_{\rm s}^{0}$) reported in Ref. \cite{nphys3104}, which are given in Appendix \ref{app:sim_details}. We find that by choosing an appropriate set of experimental parameters, the HR can be observed with well-resolved wavelengths and frequencies. 

\begin{figure*}[htb]
\centering
\includegraphics[width=6.7in]{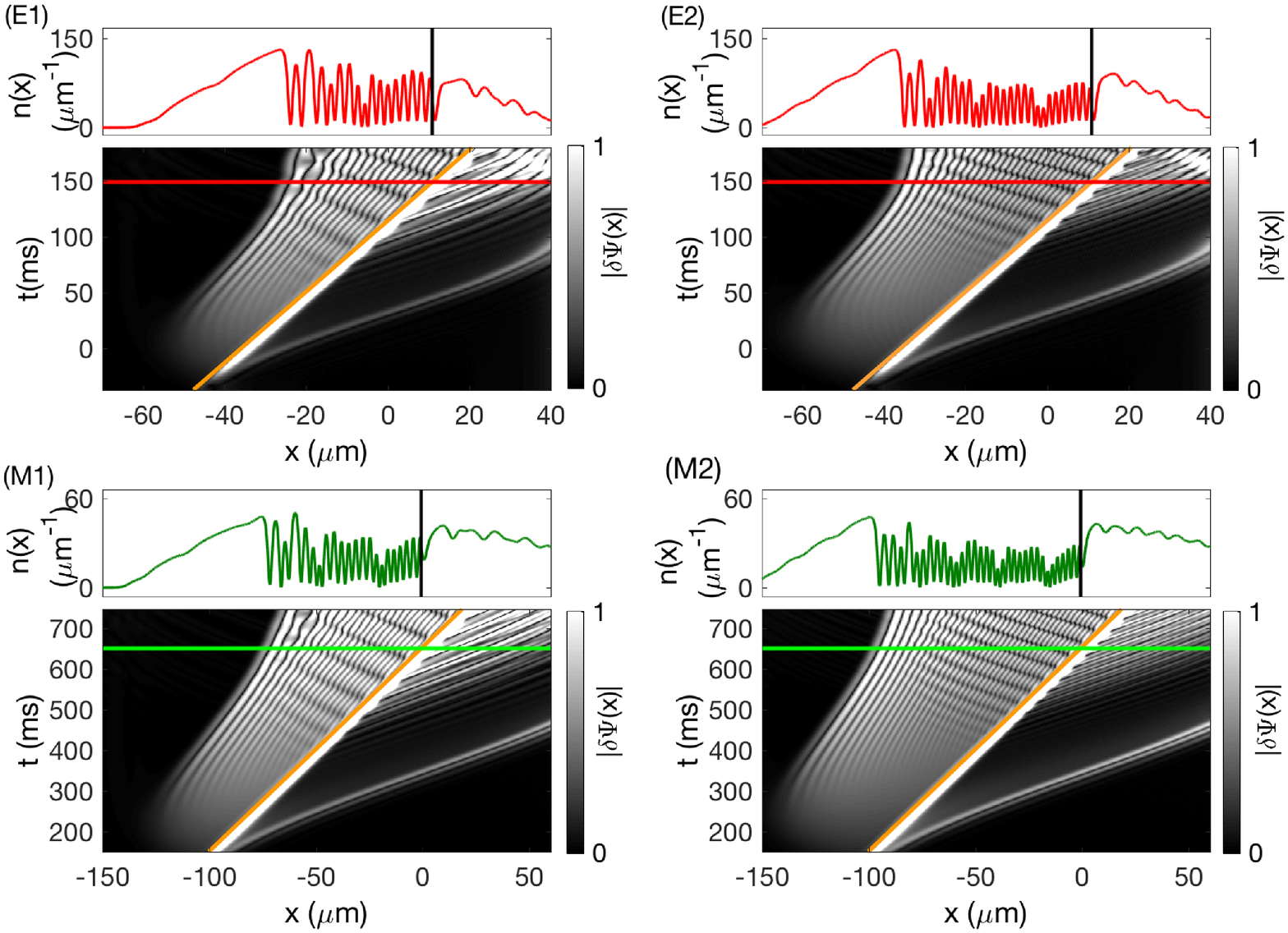}
\caption{\label{fig:evolution_modified} \textcolor{black}{Time evolution for modified parameter regimes, characterized by modified trapping frequency $\omega_x=\gamma\omega^0_x$, step speed $v_s=1.5 \gamma^{1/3}v_{\rm s}^0$, and step strength $U_{\rm s}$. Note that $\omega^0_x$ and $v_{s}^0$ are the reference values taken from \protect\cite{nphys3104}, $\omega^0_x= (2\pi) \times 4.5$ Hz, and $v_{\rm s}^0=$ 0.21 mm/s. Modified regimes: (E1) $\gamma=1$, $U_{\rm s}/k=6$ nK and $v_s=$ 1.5$v_s^0$; (E2) $\gamma=1$, $U_{\rm s}/k=9$ nK, and $v_{\rm s}=$ 1.5$v_{\rm s}^0$; (M1) $\gamma=1/4$, $U_{s}/k=\gamma^{2/3}\times 6$ nK, $v_s=$ 1.5$\gamma^{1/3}v_s^0$; (M2) $\gamma=1/4$, $U_{\rm s}/k=\gamma^{2/3}\times 9$ nK, $v_s=$ 1.5$\gamma^{1/3}v_s^0$. Bottom: time evolution of $|\delta \Psi(x,t)|$; top: density profile $n(x)$ at times indicated by the horizontal red (gray) or green (light gray) line in the lower panel. Note that $|\delta \Psi(x,t)|$ is multiplied by 10 for $x > x_{\mathrm{BH}}$, where $x_{\mathrm{BH}}$ is indicated by the diagonal orange lines.
}}
\end{figure*}

Figure~\ref{fig:evolution_modified} shows four representative cases for our investigation: E1, E2, M1, and M2. Regimes E1 and E2 use the same trapping frequency as the experimental value $\omega^{0}_x$, but adopt a greater step speed $v_s=$ 1.5$v_s^0$; case E1 uses the same step strength as Fig.~\ref{fig:stimulated_HR}(b) in the paper, $U_{\rm s}/k=6$ nK; case E2 has a greater step strength, $U_{\rm s}/k=9$ nK. Note that $\omega^0_x$ and $v_{s}^0$ are the reference values taken from \cite{nphys3104}, $\omega^0_x= (2\pi) \times 4.5$ Hz, and $v_{\rm s}^0=$ 0.21 mm/s. 

Regimes M1 and M2 are the cases equivalent to E1 and E2 with a modified trapping frequency, $\omega_x=(1/4)\omega^0_x$. Here, we use some scaling relations to determine the step speed $v_{\rm s}$ and depth $U_{\rm s}$ that give rise to an equivalent flow structure with the modified trapping frequency. We know that modifying $\omega_x$ changes the speed of sound $c$ [due to the change of $n(x)$] and the chemical potential $\mu$, and subsequently changes the flow structure shown in Fig.~\ref{fig:vc_supplement}. Using the Thomas-Fermi approximation \cite{pethick} for a 1D condensate in a harmonic trap, we find that $\mu \propto \omega_x^{2/3}$, and the maximal density $n_{\rm max}\propto \omega_x^{2/3}$ (i.e., $c_{\rm max}\propto \omega_x^{1/3}$). By keeping ratios $U_{\rm s}/\mu$ and $v_{\rm s}/c_{\rm max}$ fixed, we can construct an equivalent flow structure under a different trapping frequency. We define the scaling factor $\gamma=\omega_x/\omega_x^0$, and incorporate $\gamma$ into the ratios. This gives rise to the scaling relations, $U_{\rm s}=\gamma^{2/3}U_{\rm s}^0$ and $v_{\rm s}=\gamma^{1/3}v_{\rm s}^0$. Regime M1 is the modified case for E1, such that $U_{s}/k=\gamma^{2/3}\times 6$ nK, $v_s=$ 1.5$\gamma^{1/3}v_s^0$; likewise, M2 is the modified case for E2, so $U_{\rm s}/k=\gamma^{2/3}\times 9$ nK, $v_s=$ 1.5$\gamma^{1/3}v_s^0$.

Our investigation shows that a clear mode structure occurs in regimes where the background flow is sufficiently homogeneous. Then, the BdG modes can be described as WKB modes with well-characterized frequency and wavevector, as in Ref. \cite{parentani2010}.  In the experimental regime [Fig.~\ref{fig:stimulated_HR}(b)], $\psi_{\rm p}$ has the longest wavelength, and is comparable to the width of the BH cavity, $L$ (the distance between the BH and WH). We find that the mode structure is improved when reducing the wavelength of the partner mode $\psi_{\rm p}$, relative to $L$. 

To control the wavelength of $\psi_{\rm p}$, one can refer to the BCR mechanism and the stimulated Hawking effect, and use the dispersion relation shown in Fig.~\ref{fig:STFT_wt}(a) in the paper. Overall, the wavelength of the p-mode decreases with increasing step speed, $v_{\rm s}$. According to the dispersion relation, the BCR is the zero frequency mode in the WH frame, $\Delta\omega(\Delta k_{\rm BCR})=0$. Increasing $v_s$ increases the flow speed inside the supersonic cavity, which lowers (raises) the positive-$k$ (negative-$k$) branch of the dispersion curve $\Delta\omega(\Delta k)$, and displaces the intersection $\Delta\omega(\Delta k)=0$ to a larger $\Delta k$ value. This further increases the frequency $|\Delta\omega|$ of the Hawking pair, which is proportional to $k_{\rm BCR}$, and displaces the root of the dispersion curves for $\psi_{\rm HR}$ and $\psi_{\rm p}$ to greater $|\Delta k|$ [see Fig.~\ref{fig:STFT_wt}(a)]. In regime E1, we increase $v_{\rm s}$ by 50\% over the experimental value. This decreases the p-mode wavelength relative to the cavity length, $L$, and the corresponding HR appears more periodic. We further extend $L$, by increasing step depth, $U_s$.  Regime E2 in Fig.~\ref{fig:evolution_modified} corresponds to the case with a greater step depth, in which the number of oscillations of $\psi_{\rm p}$ doubles.

Regimes M1 and M2 adopt a smaller trapping frequency, $\omega_x=(1/4)\omega^0_x$. Reducing $\omega_{x}$ increases the size of a BEC, and extends the flow structure, by which excitation modes can be more easily observed and resolved in the laboratory. We can see that cases M1 and M2 have clear mode structures as in E1 and E2, with approximately twice the cavity length. Note that regime M2 is reported in the paper, along with a mode analysis using the spatial and temporal WFTs.

\subsection{Growth of the BCR mode in the M2 regime}
\label{app:growth}
In the paper, we found that in the experimental regime the standing-wave amplitude $n_k$ inside the cavity (which later proved to be the BCR) grows in proportion to the square of the background density $n_{\rm bf}$, $n_k \propto n_{\rm bf}^2$. We use the BCR mechanism to interpret this relationship. If it is indeed the underlying mechanism that occurs in the step-sweeping experiment, the same growth relationship should be found in other parameter regimes.
\begin{figure}[!htb]
\includegraphics[width=3.3in]{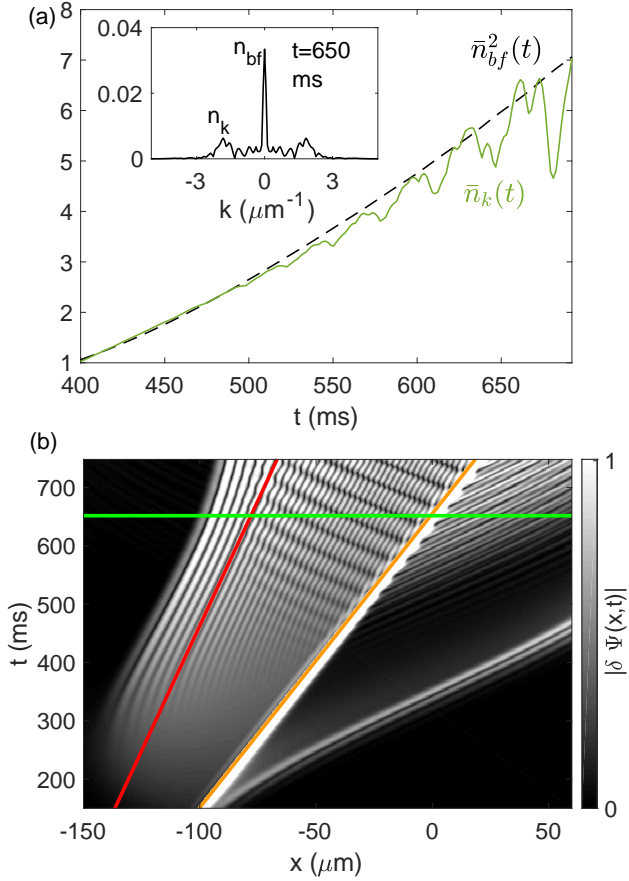}
\caption{\label{fig:M2_growth} Growth of the standing wave in the M2 regime. Panel (a) simulated growth of the standing-wave pattern in the supersonic region. Solid green: normalized standing-wave amplitude $\bar{n}_k(t)$, $\bar{n}_{k}(t)=n_{k}(t)/n_{k}(0)$.
Dashed black: the square of background density, $\bar{n}_{\rm bf}(t)$, scaled to match the final standing-wave amplitude, $\bar{n}_{\rm bf}^2(t)=n_{\rm bf}^2(t)[\bar{n}_{k}(t_f)/n_{\rm bf}^2(t_f)]$. Panel (b): time evolution of $|\delta \Psi(x,t)|$, from which we select a position nearby the WH, denoted by the diagonal red (left) line, to monitor the mode growth. The growth of $n_{\rm bf}$ and $n_{k}$ is determined from a spatial WFT of $n(x)$ with window width $D= 20$ $\mu$m at the position indicated by the diagonal red (left) line in (b). Inset shows the windowed spectrum at $\text{t=650 ms}$.}
\end{figure}

In Fig.~\ref{fig:M2_growth}, we monitor the growth of the standing wave at a position nearby the WH [indicated by the diagonal red (left) line in Fig.~\ref{fig:M2_growth}(b)] . We find that the growth of the standing wave $n_k$ [solid green curve in (a)] matches that of the background flow $n_{\rm bf}^2$ [dashed black curve in Fig.~\ref{fig:M2_growth}(a)], which is consistent with the observation in the experimental regime. For both regimes, the relationship $n_k \propto n_{\rm bf}^2$ implies that the BCR mechanism along with the increasing background density gives rise to the mode growth inside the cavity, rather than the black-hole lasing effect. Note that the p-mode propagates to the position indicated the diagonal red (left) line at $t\sim 500$ ms, which causes some small oscillations on the growth plot $n_k(t)$.

\section{Bogoliubov-de Gennes mode analysis: asymptotic modes, local dispersion relations, and the thermal prediction}
\label{app:BdG}
Here, we present the Bogoliubov--de Gennes (BdG) equations and the relevant calculations discussed in the paper. In Appendix~\ref{app:WKB}, we introduce the standard BdG formalism, and an asymptotic method (WKB) to describe modes on a slowly varying background. In Appendix~\ref{app:dispersion}, we use the dispersion relation to determine spectral properties of the modes. In Appendix~\ref{app:temperature}, we compare the mode amplitudes of the Hawking pair with 
the thermal prediction using the flow profile at the BH.

\subsection{BdG equations and asymptotic BdG modes}
\label{app:WKB}

Here, we summarize the BdG formulation presented in \cite{pethick, parentani2010}. The BdG equations can be obtained by the linearization of the condensate wavefunction:

\begin{eqnarray}
\label{eq:psi}
\Psi(x,t)=\Psi_0(x,t)+\psi(x,t)
\end{eqnarray}
where $\Psi_0(x,t)$ corresponds to a stationary condensate, $\Psi_0(x,t)=\sqrt{n(x)}e^{-i \mu t}$, and $\psi(x,t)$ corresponds to a deviation to the background condensate, which can be expressed as
\begin{eqnarray}
\label{eq:BdG1_app}
\psi(x,t) &=& e^{-i \mu t}\left( u(x) e^{-i\omega t} +v^{*}(x) e^{i\omega t}\right),
\end{eqnarray} 
where $u(x)$ and $v^{*}(x)$ satisfy the BdG equations:
\begin{eqnarray}
\label{eq:BdG2}
\left [\hbar\omega+\frac{\hbar^2}{2m}\frac{d^2 }{d x^2}-V(x)-2gn(x)+\mu \right ]u(x)&=&gn(x)v(x),\nonumber\\ 
\left [-\hbar\omega+\frac{\hbar^2}{2m}\frac{d^2 }{d x^2}-V(x)-2gn(x)+\mu \right ]v(x)&=&gn(x)u(x).\nonumber\\
\end{eqnarray} 
For a homogeneous system, BdG modes can be expressed as plane waves
\begin{eqnarray}
\label{eq:uv_plane}
u({x})=u_k \frac{e^{ikx}}{\sqrt{2\pi}} ,\,\,\, v({x})=v_k\frac{e^{ikx}}{\sqrt{2\pi}},
\end{eqnarray} 
where the normalized mode amplitudes $u_k$ and $v_k$ are 
\begin{eqnarray}
\label{eq:uv_D}
u_k=\frac{1}{\sqrt{1-D_k^2}} ,\,\,\, v_k=\frac{D_k}{\sqrt{1-D_k^2}},
\end{eqnarray} 
where $D_k$ gives the ratio between $v_k$ and $u_k$, and is determined by the speed of sound $c=\sqrt{gn/m}$,
\begin{eqnarray}
\label{eq:D}
D_k=\frac{1}{mc^2}\left[\sqrt{\hbar^2c^2k^2+\left(\frac{\hbar^2 k^2}{2m}\right)^2}-\frac{\hbar^2 k^2}{2m}-mc^2\right].\nonumber\\
\end{eqnarray} 
This leads to the dispersion relation
\begin{eqnarray}
\label{eq:dispersion_supplement}
\omega(k)^2 &=& c^2k^2+\frac{\hbar^2 k^4}{4m^2}.
\end{eqnarray}
Note that here we use $\omega$ and $k$ to indicate the relative frequency $\Delta \omega$ and wavevector $\Delta k$ adopted in the paper. Using Eq.~\ref{eq:dispersion_supplement}, the BdG modes in the wavevector ($k$) representation can be converted to the frequency ($\omega$) representation, such that $u_{\omega}=u_k/\sqrt{d\omega/dk}$ and $v_{\omega}=v_k/\sqrt{d\omega/dk}$.

Suppose the background condensate is inhomogeneous but varies smoothly in space, the BdG modes can be approximated by the WKB method as described in \cite{parentani2010}. The WKB-BdG modes in the $\omega$-representation are 
\begin{eqnarray}
\label{eq:uv_WKB}
u_{\omega}({x})=\sqrt{\frac{\partial k_{\omega}(x)}{\partial\omega}}\frac{1}{\sqrt{1-D_{k_{\omega}(x)}^2}} \frac{e^{i\int^x k_{\omega}(x')dx'}}{\sqrt{2\pi}}, \nonumber\\
v_{\omega}({x})=\sqrt{\frac{\partial k_{\omega}(x)}{\partial\omega}}\frac{D_{k_{\omega}(x)}}{\sqrt{1-D_{k_{\omega}(x)}^2}} \frac{e^{i\int^x k_{\omega}(x')dx'}}{\sqrt{2\pi}}, 
\end{eqnarray} 
where $k_{\omega}(x)$ is determined by the local dispersion relation using the local sound speed $c(x)$
\begin{eqnarray}
\label{eq:dispersion_WKB}
\omega^2=c(x)^2k^2+\frac{\hbar^2 k^4}{4m^2}.
\end{eqnarray} 
\begin{figure*}[htb]
\includegraphics[width=7in]{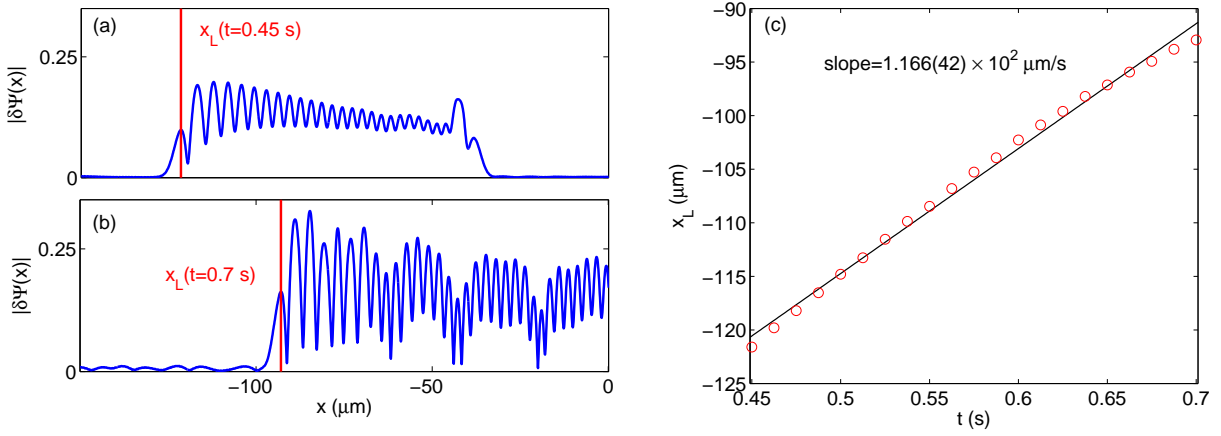}
\caption{Determination of the WH velocity. The left edge of $|\delta\Psi(x)|$, $x_{\rm L}$, is measured at various times, as indicated by the vertical red lines in (a) and (b), and the red circles in (c). The speed of the left edge, $v_{\rm L}$, is determined by a linear fit on $x_{\rm L}(t)$, as indicated by the black line in (c), $v_{\rm L}\sim$ 0.117 mm/s.}
\label{fig:v_WH}
\end{figure*}
\subsection{Dispersion relations for the BCR mechanism and stimulated pair production}
\label{app:dispersion}

\begin{figure*}[!ht]
\centering
\includegraphics[width=6.8in]{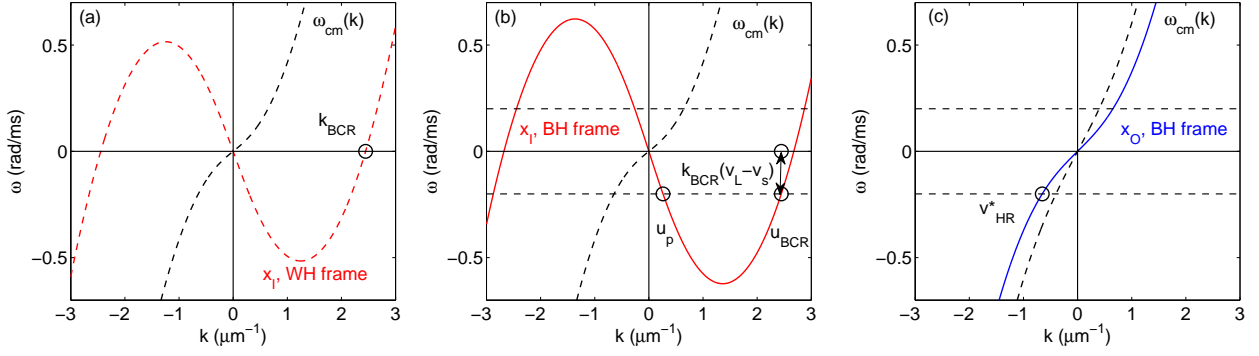}
\caption{\label{fig:dispersion} 
Dispersion relations. (a) Dashed red (gray): the dispersion relation calculated at $x_{\rm I}$ in the WH frame; (b) solid red: the dispersion relation calculated at $x_{\rm I}$ in the BH frame; (c) solid blue: the dispersion relation calculated at $x_{\rm O}$ in the BH frame. Dashed black curve: dispersion relations in the comoving frame of the condensate. The BCR is the zero-frequency mode in the WH frame (a), from which its wavevector $k_{\rm BCR}$ is determined. Transforming the BCR to the BH frames gives rise to the frequency of the HR and partner modes (indicated by horizontal lines) as shown in (b) and (c). }
\end{figure*}

Here, we use the local dispersion relations to determine the wavevectors and frequencies of the three BdG modes: $\psi_{\rm BCR}$, $\psi_{\rm HR}$, and $\psi_{\rm p}$. First, we transform the local dispersion relation from the comoving frame of the condensate [Eq.~\ref{eq:dispersion_WKB}, denoted by $\omega_{\rm cm}(k)$] to an observer frame in which the condensate has nonzero flow velocity
\begin{eqnarray}
\label{eq:dispersion_WKB_v}
\omega &=&\sqrt{c(x)^2k^2+\frac{\hbar^2 k^4}{4m^2}}+v_{\rm bf, o}(x)k,
\end{eqnarray}
where $v_{\rm bf,o}(x)$ is the local velocity of the condensate with respect to the ``observer" frame in which the frequency is defined. Then, we select two points of observation: $x_{\rm I}$ inside the BH cavity, and $x_{\rm O}$ outside. The local speed of sound and the flow velocity can be evaluated by the spatial WFT of $\Psi(x)$, in which the dominant peak location gives the wavevector of background flow in the laboratory frame $k_{\rm bf}(x)$, and its peak value gives the local density $n_{\rm bf}(x)=|\Psi(k_{\rm bf},x)|^2$. The flow velocity in the BH frame is $v_{\rm bf,BH}(x)=\hbar k_{\rm bf}(x)/m-v_{\rm s}$. 
The WH is defined by the point where $v(x)+c(x)=0$. It is formed in the small transition region that connects the accelerated flow and the $k\sim0$ region on the left [see Fig.~\ref{fig:exp}(b)], which can be identified by the left edge of $|\delta\Psi(x,t)|$ in Fig. \ref{fig:v_WH}, $x_{\rm L}(t)$. Thus, we approximate the speed of WH by that of the left edge, $v_{\rm L}$, as shown in Fig. \ref{fig:v_WH}. The flow velocity in the WH frame is given by $v_{\rm bf,WH}(x)=\hbar k_{\rm bf}(x)/m-v_{\rm L}$. The change of flow velocity from the WH frame to the BH frame is equal to the velocity difference $\Delta v$ between the two horizons, $v_{\rm bf,BH}-v_{\rm bf,WH}=-\Delta v=-(v_{\rm s}-v_{\rm L})$.

According to the BCR mechanism, the BCR mode has zero frequency in the WH frame, $\omega(k_{\rm BCR})=0$. We can predict the value of $k_{\rm BCR}$ by the local dispersion relation at $x_{\rm I}$ using $c(x_{\rm I})$ and $v_{\rm bf,WH}(x_{\rm I})$, as shown in Fig.~\ref{fig:dispersion}(a). This further determines the frequency of the stimulated pair production [see Figs.~\ref{fig:dispersion}(b) and (c)]; the HR and p modes have the same frequency as that of the BCR mode in the BH frame, given by $\omega=-\Delta v k_{\rm BCR}$. This frequency intersects the dispersion relation evaluated at $x_{\rm I}$ to determine $k_{\rm p}$, and the one evaluated at $x_{\rm O}$ to determine $k_{\rm HR}$, as indicated in Figs.~\ref{fig:dispersion}(b) and (c).

\subsection{Stimulated pair creation with the thermal prediction}
\label{app:temperature}
The Hawking temperature can be estimated by measuring the amplitudes of the correlated HR and partner mode. The mode mixing process at the BH is expressed by
\begin{eqnarray}
\label{eq:alpha_beta}
u_{-\omega}^{\rm BCR}=\alpha_{-\omega}^{\rm p}u_{-\omega}^{\rm p}+\beta_{\omega}^{\rm HR}{\left(v_{\omega}^{\rm HR}\right)}^*,
\end{eqnarray}
where $\alpha_{-\omega}^{\rm p}$ and $\beta_{\omega}^{\rm HR}$ are the positive-norm and negative-norm amplitudes of the Hawking pair, and $u_{-\omega}^{\rm BCR}$ and $u_{-\omega}^{\rm p}$ are the $u$ component of the BCR and the partner mode, $(v_{\omega}^{\rm HR})^*$ the $v$ component of the HR mode. The ratio of the amplitudes can be calculated using the thermal prediction \cite{robertson2012,PhysRevA.80.043601}
\begin{eqnarray}
\label{eq:Hawking}
\left|\frac{\beta_{\omega}^{\rm HR}}{\alpha_{-\omega}^{\rm p}}\right|=e^{-\frac{\pi\omega}{\kappa}},
\end{eqnarray}
where $\kappa$ is the surface gravity at the BH determining the Hawking temperature $T_{\rm H}=\hbar\kappa/(2\pi k)$, 
\begin{eqnarray}
\label{eq:kappa}
\kappa=\left.\frac{d\left(v+c\right)}{dx}\right|_{x_{\rm BH}}.
\end{eqnarray}
In Fig.~\ref{fig:M20}(b) in the paper, we trace a correlated Hawking pair [indicated by the magenta (left) and cyan (right) dots] generated at $t=$ 588 ms. Using the spatial WFTs, we obtain the mode amplitudes of the pair, ${v_{\omega, {\rm FT}}^{\rm HR}}^*$ and ${u_{-\omega,{\rm FT}}^{\rm p}}$. Using Eqs.~\ref{eq:uv_WKB} and \ref{eq:dispersion_WKB}, they can be expressed in relation to $\left|\beta_{\omega}^{\rm HR}/\alpha_{-\omega}^{\rm p}\right|$ as
\begin{eqnarray}
\label{eq:Hawking_FT}
\left|\frac{{v_{\omega,{\rm FT}}^{\rm HR}}^*}{u_{-\omega,{\rm FT}}^{\rm p}}\right| & = & \left|\frac{\beta_{\omega}^{\rm HR}}{\alpha_{-\omega}^{\rm p}}\right| \sqrt{\left(1-D_{k_{\rm p}}^2\right)\left( \frac{D_{k_{\rm HR}}^2}{1-D_{k_{\rm HR}}^2}\right)} \nonumber\\
&&~\times \left| \frac{\partial \omega/\partial k|_{k_{\rm p}}}{\partial \omega/\partial k|_{k_{\rm HR}}}\right|^{1/2},
\end{eqnarray}
where $D_{k_{\rm HR}}$ (and $D_{k_{\rm p}}$) can be evaluated using Eq.~\ref{eq:D}, and $\partial \omega/\partial k|_{k=k_{\rm HR}}$ estimated using the dispersion relation.

\begin{figure}[htb]
\includegraphics[width=3.1in]{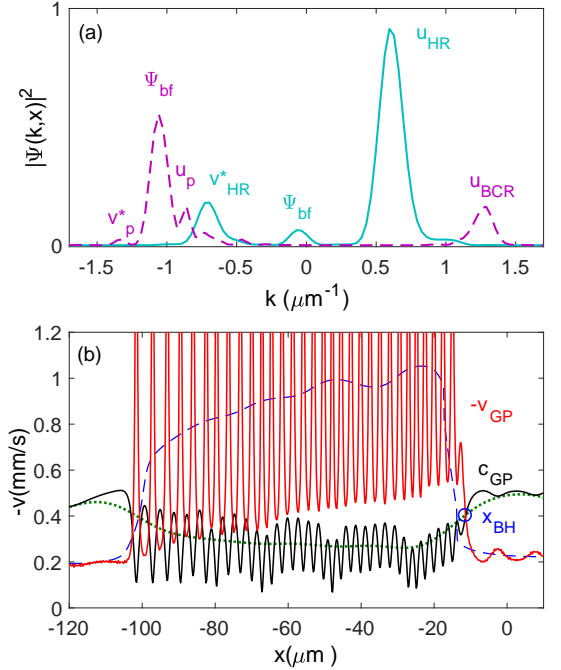}
\caption{Spatial WFT for a correlated Hawking pair, and surface gravity near the event horizon. (a) Windowed wavevector spectrum of the correlated HR (solid cyan curve) and partner modes (dashed magenta) at $t=650$ ms. (b) Flow velocity $v_{\rm GP}(x)$ and speed of sound $c_{\rm GP}(x)$ at the time at which the pair is created ($t=588$ ms). The BH is indicated by the blue circle, $x_{\rm BH}$. The flow velocity (dashed blue) and the speed of sound (dotted green) calculated from the spatial WFT are plotted for comparison. The surface gravity $\kappa$ is calculated from the speed slopes at the BH. }
\label{fig:kappa}
\end{figure}
\begin{table}[ht]
\caption{Numerical values of the relative mode amplitude and the relevant quantities in Eq.~\ref{eq:Hawking_FT}.}
\label{table:temperature}
\begin{tabular}{l c c }
\hline \hline
Quantity   & Value  & \\ \hline
$|{v_{\omega,{\rm FT}}^{\rm HR}}^*/{u_{-\omega,{\rm FT}}^{\rm p}}|$   & 0.11 &  \\
$|D_{\rm p}|$     &     0.47 (BdG) &  0.40 (FT)    \\
$|D_{\rm HR}|$  &     0.39 (BdG) &  0.44 (FT)    \\
$\partial \omega/\partial k|_{k_{\rm p}} $   &  -744 $\mu$m/ms & \\
$\partial \omega/\partial k|_{k_{\rm HR}} $  &  409 $\mu$m/ms &  \\
$\left|\beta_{\omega}^{\rm HR}/\alpha_{-\omega}^{\rm p}\right|$  & 0.21  &\\
\hline \hline
\end{tabular}
\end{table}

Figure~\ref{fig:kappa}(a) shows the windowed wavevector spectra of a correlated Hawking pair at $t=650$ ms. The numerical values of the relevant quantities in Eq.~\ref{eq:Hawking_FT} are given in Table \ref{table:temperature}. The ratio $\left|\beta_{\omega}^{\rm HR}/\alpha_{-\omega}^{\rm p}\right|\sim 0.21 \, $. To estimate how well the linear (BdG) approximation works, we measure the quantities $|D_{\rm p}|$ and $|D_{\rm HR}|$ from the wavevector spectrum (denoted by ``FT'' in Table \ref{table:temperature}), which correspond to the ratio between the $u$ and $v$ amplitudes of each mode, as indicated in Eq.~\ref{eq:uv_WKB}. They differ from the calculated (BdG) values by 13 \% for the HR mode, and 15 \% for the partner mode.

We also calculate $\left|\beta_{\omega}^{\rm HR}/\alpha_{-\omega}^{\rm p}\right|$ using the thermal prediction (Eq.~\ref{eq:Hawking}) and the surface gravity $\kappa$ from (Eq.~\ref{eq:kappa}). From the data in Fig.~\ref{fig:kappa}(b), we calculate $\kappa = 350 \,$ s$^{-1}$.
This corresponds to a Hawking temperature of $T_{\rm H}=$ 0.43 nK. The ratio $\left|\beta_{\omega}^{\rm HR}/\alpha_{-\omega}^{\rm p}\right|\sim 0.17$, as determined by the thermal prediction, differs from the WFT value ($\sim0.21$) by 24\%.

\bibliography{HawkingCondensateFinal}

\end{document}